\documentclass[10pt,twocolumn]{IEEEtran} %

\usepackage{amsmath,amssymb,amsfonts,acronym}
\usepackage{hyperref}
\usepackage{algorithmic}
\usepackage[usenames]{color}
\usepackage{psfrag}
\usepackage{bm}
\usepackage{graphicx}

\usepackage[permil]{overpic}
\usepackage{pict2e}
\usepackage{pgfplots}
\usepackage{pgfplotstable}
\pgfplotsset{compat=newest}

 \usetikzlibrary{plotmarks}
  \usetikzlibrary{arrows.meta}
  \usepgfplotslibrary{patchplots}
  \usepackage{grffile}
  \newcommand{\figurewifthlatex}{0.8}
\usepackage{tikz}

 \usepackage{subcaption}
 \usepackage{cite}
 \usepackage[capitalise]{cleveref}
\Crefname{equation}{eq:\!}{Eqs.\!}
\Crefname{figure}{Fig.\!}{Figs.\!}
\Crefname{tabular}{Tab.\!}{Tabs.\!}
\Crefname{section}{Section\!}{Sections.\!}
\title{Reconfigurable Intelligent Surfaces for Localization: Position and Orientation Error Bounds}
\author{Ahmed Elzanaty,~\IEEEmembership{Member,~IEEE,} Anna Guerra,~\IEEEmembership{Member,~IEEE,} \\ Francesco Guidi,~\IEEEmembership{Member,~IEEE,} 
and Mohamed-Slim Alouini,~\IEEEmembership{Fellow,~IEEE,}
\thanks{Ahmed Elzanaty and Mohamed-Slim Alouini are with  King Abdullah University of Science and Technology (KAUST), Thuwal 23955-6900, Saudi Arabia, e-mail: \{ahmed.elzanaty, slim.alouini\}@kaust.edu.sa.
Anna Guerra is with the WiLAB, University of Bologna, Italy, email: anna.guerra3@unibo.it.
Francesco Guidi is with CNR-IEIIT, Italy, e-mail: francesco.guidi@ieiit.cnr.it.
}
}
\begin{document}
\pgfplotsset{compat=1.14}

\newcommand{\Nb} {N_\mathsf{B}}
\newcommand{\Nm} {N_\mathsf{M}}
\newcommand{\Nl} {N_\mathsf{R}}
\newcommand{\Ns} {N}

\newcommand{\Hm} {\mathbf{H}}
\newcommand{\y} {\mathbf{y}}
\newcommand{\x} {\mathbf{x}}
\newcommand{\n} {\boldsymbol{\omega}}
\newcommand{\Hbm} {\mathbf{H}_\mathsf{BM}}
\newcommand{\Hbl} {\mathbf{H}_\mathsf{BR}}
\newcommand{\Hlm} {\mathbf{H}_\mathsf{RM}}
\newcommand{\bmOmega} {\bm{\Omega}}
\newcommand{\alphar} {\bm{\alpha}_\mathsf{r}}
\newcommand{\alphat} {\bm{\alpha}_\mathsf{t}}

\newcommand{\rhobm} {\rho_\mathsf{BM}}
\newcommand{\rhobl} {\rho_\mathsf{BR}}
\newcommand{\rholm} {\rho_\mathsf{RM}}
\newcommand{\rhoblm} {\rho_\mathsf{BRM}}

\newcommand{\tbm} {\tau_{bm}}
\newcommand{\tbl} {\tau_{br}}
\newcommand{\tlm} {\tau_{rm}}
\newcommand{\tdm} {\tau_{sm}}

\newcommand{\taubm} {\tau_\mathsf{BM}}
\newcommand{\taubl} {\tau_\mathsf{BR}}
\newcommand{\taulm} {\tau_\mathsf{RM}}
\newcommand{\taudm} {\tau_\mathsf{DM}}

\newcommand{\dbmo} {d_\mathsf{BM}}
\newcommand{\ddmo} {d_\mathsf{SM}}
\newcommand{\dblo} {d_\mathsf{BR}}
\newcommand{\dlmo} {d_\mathsf{RM}}

\newcommand{\xilm} {\xi_{\mathsf{RM}}}
\newcommand{\etal} {\eta_r}
\newcommand{\etam} {\eta_m}

\newcommand{\xibm} {\xi_{\mathsf{BM}}}

\newcommand{\chibm}{\chi_{\mathsf{B}\mathsf{M}}}
\newcommand{\chiblm}{\chi_{\mathsf{B}\mathsf{R}\mathsf{M}}}

\newcommand{\phibm} {\phi_\mathsf{BM}}
\newcommand{\phibl} {\phi_\mathsf{BR}}
\newcommand{\philm} {\phi_\mathsf{RM}}
\newcommand{\phiL} {\phi_{\mathsf{R}}}
\newcommand{\phidm} {\phi_\mathsf{SM}}

\newcommand{\thetadm} {\theta_\mathsf{SM}}
\newcommand{\thetabm} {\theta_\mathsf{BM}}
\newcommand{\thetabl} {\theta_\mathsf{BR}}
\newcommand{\thetalm} {\theta_\mathsf{RM}}

\newcommand{\thetam} {\theta_m}
\newcommand{\thetab} {\theta_b}
\newcommand{\thetal} {\theta_l}

\newcommand{\xms} {x_{\mathsf{M}}}
\newcommand{\yms} {y_{\mathsf{M}}}
\newcommand{\zms} {z_{\mathsf{M}}}

\newcommand{\phim} {\phi_m}
\newcommand{\phib} {\phi_b}
\newcommand{\phil} {\phi_r}

\newcommand{\thetaL} {\theta_{\mathsf{R}}}

\newcommand{\dm} {d_{m}}
\newcommand{\dl} {d_{r}}
\newcommand{\db} {d_{b}}
\newcommand{\dbm} {d_{bm}}
\newcommand{\ddm} {d_{sm}}
\newcommand{\dbl} {d_{br}}
\newcommand{\dlm} {d_{rm}}
\newcommand{\fn} {f_n}

\newcommand{\xb}{x_b}
\newcommand{\yb}{y_b}
\newcommand{\nb}{w}
\newcommand{\zb}{z_b}

\newcommand{\xm}{x_m}
\newcommand{\ym}{y_m}
\newcommand{\zm}{z_m}

\newcommand{\xd}{x_s}
\newcommand{\yd}{y_s}
\newcommand{\zd}{z_s}

\newcommand{\Cbrm}{C_{brm}}
\newcommand{\BCr}{\bar{C}_{r}}
\newcommand{\BCk}{\bar{C}_{k}}
\newcommand{\Cbkm}{C_{bkm}}
\newcommand{\betam}{\beta_{m}}
\newcommand{\fc}{f_{0}}
\newcommand{\thetar}{\theta_r}
\newcommand{\thetak}{\theta_k}

\newcommand{\xdi}{x_d^{(0)}}
\newcommand{\ydi}{y_d^{(0)}}
\newcommand{\zdi}{z_d^{(0)}}

\newcommand{\xmi}{x_m^{(0)}}
\newcommand{\ymi}{y_m^{(0)}}
\newcommand{\zmi}{z_m^{(0)}}

\newcommand{\cosam}{c_{\alpha_{\mathsf{M}}}}
\newcommand{\sinam}{s_{\alpha_{\mathsf{M}}}}
\newcommand{\cosbm}{c_{\beta_{\mathsf{M}}}}
\newcommand{\sinbm}{s_{\beta_{\mathsf{M}}}}

\newcommand{\cosgm}{c_{\gamma_{\mathsf{M}}}}
\newcommand{\singm}{s_{\gamma_{\mathsf{M}}}}

\newcommand{\xl}{x_r}
\newcommand{\yl}{y_r}
\newcommand{\zl}{z_r}
\newcommand{\godm}{\mathsf{G}^{(1)}_{sm}}
\newcommand{\gobm}{\mathsf{G}^{(1)}_{bm}}
\newcommand{\gobl}{\mathsf{G}^{(1)}_{bl}}
\newcommand{\gtbm}{\mathsf{G}^{(2)}_{bm}}
\newcommand{\gtdm}{\mathsf{G}^{(2)}_{sm}}
\newcommand{\gtBM}{G_{\mathsf{B,M}}^{(b,m)}}
\newcommand{\gtDM}{G_{\mathsf{D,M}}^{(d,m)}}

\newcommand{\gtbl}{\mathsf{G}^{(2)}_{br}}
\newcommand{\gtlm}{\mathsf{G}^{(2)}_{rm}}
\newcommand{\gtLM}{G_{\mathsf{R,M}}^{(r,m)}}

\newcommand{\gtBL}{G_{\mathsf{B,R}}^{(b,r)}}

\newcommand{\xmo}{x_{\mathsf{M}}}
\newcommand{\ymo}{y_{\mathsf{M}}} 
\newcommand{\zmo}{z_{\mathsf{M}}} 
\newcommand{\pmo}{\mathbf{p}_{\mathsf{M}}}

\newcommand{\pdo}{\mathbf{p}_{\mathsf{D}}}
\newcommand{\ado}{\boldsymbol{\Phi}_{\mathsf{D}}}
\newcommand{\amo}{\boldsymbol{\phi}_{\mathsf{M}}}
\newcommand{\abo}{\boldsymbol{\Phi}_{\mathsf{B}}}
\newcommand{\alo}{\boldsymbol{\Phi}_{\mathsf{R}}}

\newcommand{\am}{{a}_{\mathsf{M}}}

\newcommand{\xbo}{x_{\mathsf{B}}}
\newcommand{\ybo}{y_{\mathsf{B}}} 
\newcommand{\zbo}{z_{\mathsf{B}}} 
\newcommand{\pbo}{\mathbf{p}_{\mathsf{B}}}

\newcommand{\xlo}{x_{\mathsf{R}}}
\newcommand{\ylo}{y_{\mathsf{R}}} 
\newcommand{\zlo}{z_{\mathsf{R}}} 
\newcommand{\plo}{\mathbf{p}_{\mathsf{R}}}

\newcommand{\hbm}{h_{bm}}
\newcommand{\hbl}{h_{br}}
\newcommand{\hlm}{h_{rm}}
\newcommand{\omegal}{\omega_{r}}

\newcommand{\xbs}{{x}_{\mathsf{B}}}
\newcommand{\ybs}{{y}_{\mathsf{B}}}
\newcommand{\zbs}{{z}_{\mathsf{B}}}
\newcommand{\xds}{{x}_{\mathsf{D}}}
\newcommand{\yds}{{y}_{\mathsf{D}}}
\newcommand{\zds}{{z}_{\mathsf{D}}}
\newcommand{\xls}{{x}_{\mathsf{R}}}
\newcommand{\yls}{{y}_{\mathsf{R}}}
\newcommand{\zls}{{z}_{\mathsf{R}}}
\newcommand{\dant}{{d}_{\mathsf{ant}}}
\newcommand{\muBM} {\mu_{b,\mathsf{BM}}[n]}
\newcommand{\muBLM} {\mu_{b,\mathsf{BRM}}[n]}
\newcommand{\muBMc} {\mu^{*}_\mathsf{BM}[n]}
\newcommand{\muBLMc} {\mu^{*}_\mathsf{BRM}[n]}

\newcommand{\mubm} {\mu_{bm}[n]}
\newcommand{\cmubmp} {\mu_\mathsf{bm'}^*[n]}

\newcommand{\mublm} {\mu_{brm}[n]}
\newcommand{\mubmc} {\mu^{*}_\mathsf{bm}[n]}
\newcommand{\mublmc} {\mu^{*}_\mathsf{brm}[n]}

\newcommand{\alphamo}{\alpha_{\mathsf{M}}}
\newcommand{\betamo}{\beta_{\mathsf{M}}}
\newcommand{\gammamo}{\gamma_{\mathsf{M}}}

\newcommand{\Phim}{\Phi_{\mathsf M}}

\newcommand{\PEB}{\mathsf{PEB}}
\newcommand{\OEB}{\mathsf{OEB}}
\acrodef{5G}{fifth generation mobile networks}
\acrodef{6G}{sixth generation mobile networks}
\acrodef{BS}[gNB]{next generation NodeB base station}
\acrodef{CRLB}{Cramer-Rao lower bound}
\acrodef{EM}{electromagnetic}
\acrodef{FIM}{Fisher information matrix}
\acrodef{KKT}{Karush–Kuhn–Tucker}
\acrodef{KPI}{key performance indicator}
\acrodef{LOS}{line-of-sight}
\acrodef{CSI}{channel state information}
\acrodef{EMF}{electromagnetic fields}
\acrodef{LIS}{large intelligent surface}
\acrodef{mm-wave}{millimeter-waves}
\acrodef{MLE}{maximum likelihood estimator}
\acrodef{MS}[UE]{user equipment}
\acrodef{NLOS}{non-line-of-sight}
\acrodef{OFDM}{orthogonal frequency-division multiplexing}
\acrodef{RIS}{reconfigurable intelligent surface}
\acrodef{RSS}{received signal strength}
\acrodef{SDS}{software defined surface}
\acrodef{SRE}{smart radio environment}
\acrodef{SNR}{signal-to-noise ratio}
\acrodef{TOA}{time-of-arrival}
\acrodef{AOA}{angle-of-arrival}
\acrodef{RSSI}{received signal strength indicator}
\acrodef{CRLB}{Cramér-Rao lower bound}
\acrodef{GDOP}{geometric dilution of precision}
\acrodef{PEB}{position error bound}
\acrodef{OEB}{orientation error bound}
\acrodef{RMSE}{root mean square error}
\acrodef{UAV}{unmanned aerial vehicle}

\maketitle

\vspace{-1.5 cm}
\begin{abstract}
Next-generation cellular networks will witness the creation of \acp{SRE}, where walls and objects can be coated with \acp{RIS} to strengthen the communication and localization coverage by controlling the reflected multipath. In fact, \acp{RIS} have been recently introduced not only to overcome communication blockages due to obstacles but also for high-precision localization  of mobile users in GPS denied environments, e.g., indoors.
Towards this vision, this paper presents the localization performance limits for communication scenarios where a single \ac{BS},  equipped with multiple-antennas, infers the position and the orientation of a \ac{MS} in a \ac{RIS}-assisted \ac{SRE}. We  consider a  signal model that is valid also for near-field propagation conditions, as the usually adopted far-field assumption does not always hold, especially for large \acp{RIS}. 
%
For the considered scenario, we derive the  \ac{CRLB} for assessing the ultimate localization and orientation performance of synchronous and asynchronous signaling schemes. In addition, we propose a closed-form \ac{RIS} phase profile that well suits joint communication and localization.
We perform extensive numerical results to assess the performance of our scheme for various localization scenarios and \ac{RIS} phase design.
Numerical results show that the proposed scheme can achieve remarkable performance, even in asynchronous signaling and that the proposed phase design approaches the numerical optimal phase design that minimizes the \ac{CRLB}. 
\end{abstract}
\acresetall


\bstctlcite{IEEEexample:BSTcontrol}
\IEEEpeerreviewmaketitle

\section{Introduction}
 Recently, \acp{SRE} have been conceived as a new paradigm where the traditional radio environment is turned into a smart reconfigurable space that plays an active role in transferring and processing the information \cite{di2019smart,NadEtAl:J20_2}. 
Indeed, \acp{KPI} for the next \ac{6G} promote continuous connection availability, 
strong reliability,  huge device density ($10^7$ devices per km$^2$) and air interface latency of sub-milliseconds (e.g., $10\mu$s), etc. \cite{BenDebPoo:J18,ZonEtAl:J19}.
To meet these requirements, \acp{RIS} might represent a key solution, allowing to enhance not only wireless communications but also imaging- and localization-based applications thanks to the augmented ambient awareness \cite{sarieddeen2019next,ZonEtAl:J19}. In this regard,  \acp{RIS} can aid in establishing a \ac{LOS} link between the transmitter and the receiver even in the presence of obstructions or when the received power from the direct path does not enable a robust connection \cite{TahAlrAlk:C19}. 

In analogy with software defined radios, \acp{RIS} are often referred to as \acp{SDS}, where the electromagnet response to the incident wave can be controlled by a software \cite{basar2019wireless}. The realization of such a technology might be enabled by metamaterials, which are a class of artificial materials whose physical properties, e.g., permittivity and permeability, can be engineered to exhibit some desired characteristics \cite{smith2000composite, smith2004metamaterials,engheta2006metamaterials}. 
When such metamaterials are deployed in metasurfaces, their effective parameters can be tailored to realize a desired transformation on the transmitted, received, or impinging waves \cite{Smith-2012APM,shlezinger2019dynamic,DSmith-2018TCOM,wang2019dynamic2}. With the availability of new degrees of freedom useful to improve the network performance, the environment will be no more perceived as a passive entity, but as a meaningful support for wireless communications based applications \cite{dardari2019communicating,dardari2020using,HuaHuAlexRenzo:19,NadEtAl:J20,KishkSlim:20}, e.g., energy transfer
\cite{WuZha:J20}, vehicular networks \cite{masini2020use}, \ac{UAV} communications \cite{YanMenRenzo:20}, physical layer security \cite{WuZha:J20}, cognitive radio \cite{ZhaWanPan:20}, \ac{EMF}-aware beamforming \cite{ElzanatyLucaSlim:20},  and many others \cite{NasirSlimRIS:20}. 
In this context, wireless localization with \acp{RIS} \cite{wymeersch2019radio,Juntti:19} has not yet received a large attention, albeit they represent a promising candidate for enhancing positioning and orientation estimation capabilities in next-generation cellular networks for  various \ac{6G} applications, e.g., augmented reality and self-driving cars \cite{razavizadeh2014three,moghaddam2017joint,razavizadeh20203d,masini2020use}. 
This is of great help in GPS-denied environments, and it allows to avoid the use of a dedicated infrastructure, usually made of multiple anchors. 
Indeed, the possibility to localize with antenna arrays is not new, and has been investigated in the last few years \cite{shen2010accuracy,GueGuiDar:J18,GueGuiDar:C17,GueGuiDar:C15,han2015performance,WanWuShe:J19-2}. In particular,  \ac{5G} and beyond foresee the use of \ac{mm-wave} to enable the integration of arrays with a large number of antennas (massive arrays) into small areas. By enabling such an architecture capable to realize near-pencil beam antennas, it becomes feasible not only to boost communication but also single-anchor localization capabilities at an unprecedented scale \cite{guidi2018indoor,RohEtAl:J14,GuiGueDar:J16,witrisal2016high,GuiEtAl:J17}.

Current state-of-the-art for intelligent surfaces-based localization considers studies employing \ac{RIS} either in receive mode \cite{HuRusEdf:18} or in reflection mode \cite{Juntti:19,he2019adaptive}.
When exploited in receive mode, a large intelligent surface is used to localize a user in front of it, both in near-field and far-field  \cite{HuRusEdf:17,HuRusEdf:18}.
When instead operating in reflection mode,
in \cite{zhang2020towards} it is proposed an approach exploiting the modification of the \ac{RIS} reflection coefficient such that the experienced \ac{RSS} at different points is enlarged and the localization accuracy is improved.
Differently, in \cite{Juntti:19}, authors exploit a \ac{RIS} for supporting the positioning and communication in the mm-wave frequency bands. This paper assumes that the mobile is in far-field with respect to the \ac{RIS} \cite{Juntti:19}, but this approximation is not always valid, especially when large surfaces and arrays are considered with respect to the distance. Consequently, the entailed models are no more accurate, as the mobile is not in the Fraunhofer region but in the Fresnel region, and the planar wavefront approximation does not hold.
Additionally, ignoring the spherical wavefront discard essential information regarding the location and orientation of the mobile \cite{guidi2019radio,bjornson2020power,NepBuf:J17}.

\begin{figure}[t]
	\psfrag{RIS}[c][c][0.8]{\quad \quad\quad \quad\quad \quad\quad \quad RIS ($\bmOmega$)}
	\psfrag{B}[lc][lc][0.8]{\acs{BS}/AP (Receiver)}
	\psfrag{MS}[rc][rc][0.8]{\acs{MS} (Transmitter)}
	\psfrag{h}[c][c][0.8]{$\Hbm$}
	\psfrag{hh}[c][c][0.8]{$\Hlm$}
	\psfrag{hhh}[c][c][0.8]{$\Hbl$}
	\centering
	\includegraphics[width=0.9\linewidth,draft=false]{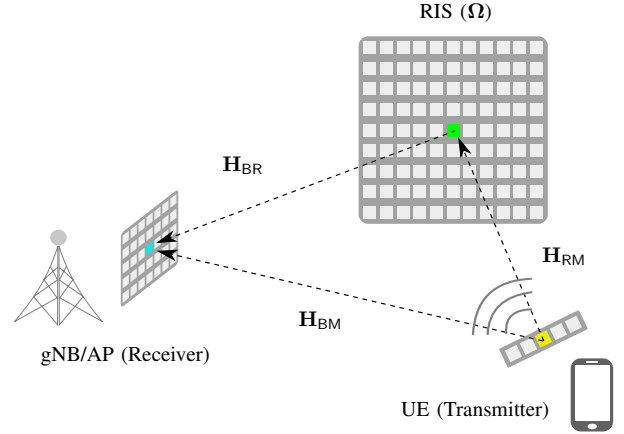}
	\caption{Pictorial representation of the considered RIS-aided positioning scenario. 
	}
	\label{fig:scenario}
\end{figure}

 
To the best of authors' knowledge, no paper has considered a general model accounting for $3$D \ac{RIS}-assisted localization and orientation estimation in near-field, as current papers for near-field positioning only refer to the adoption of a large intelligent surface in receive mode and not as a mean for controlling the multipath. To this purpose, in this paper we consider the localization scenario depicted in Fig.~\ref{fig:scenario}, where we propose an ad-hoc model for joint communication and localization, accounting for the incident spherical wavefront. Indeed, in \ac{SRE}, the \ac{BS} augments its environment awareness, as it allows also to achieve a knowledge of the environment in terms of inferring the location in the 3D space and the orientation (i.e., roll, pitch, and yaw) of the \ac{MS}. In this context, we derive the \ac{CRLB} to investigate the ultimate positioning and orientation estimation performance in the presence of the \ac{RIS}. Then, we analyze the \ac{GDOP} to evaluate the impact of the geometry on the \ac{MS} localization. Finally, we derive a suboptimal phase design for the \ac{RIS} in closed-form to enhance both the localization and communication performance by maximizing the \ac{SNR}.

The main contributions can be summarized as follows.

\begin{itemize}
    \item We propose an architecture where the \ac{RIS} is used to assist mobile localization (position and orientation estimation) at the \ac{BS}, which can provide surveillance solutions or assist the communication process, while in the literature the \ac{MS} estimates its own position.
    \item We consider that the \ac{BS}, \ac{RIS}, and the \ac{MS} are equipped with multiple-antennas with arbitrary array configurations and geometry including planar arrays that will be adopted for beyond \ac{5G} systems, especially the \ac{RIS}, allowing 3D beamforming in both the azimuth and elevation, while most of the literature considers linear arrays with 2D beamforming that significantly simplifies the analysis to the conventional steering vectors.
    \item We consider a general model valid for both near- and far-field localization and attitude (i.e., orientation) estimation in $3$D space, unlike the literature that either imposes the far-field assumption or it considers simplified $2$D geometry.
    \item Differently from the state-of-the-art that analyzes synchronous systems, we consider two general signaling schemes (i.e., synchronous and asynchronous) and compare their localization error performance. 
    \item We derive the ultimate bound on the localization performance in terms of the \ac{CRLB}. Furthermore, in order to get more insights on the effect of the geometry (e.g., the locations and orientations of the \ac{BS}, \ac{RIS}, and \ac{MS}) on the localization performance, we consider the \ac{GDOP} metric.
     \item We propose a closed-form \ac{RIS} phase design, that accounts for the spherical wavefront, and we compare it to other strategies accounting also for the presence of quantization errors; 
    \item  We perform extensive simulations and numerical results that provide insights into the problem and shed light on the benefits offered by the adoption of the \ac{RIS} in terms of localization performance.
\end{itemize}
The rest of the manuscript is organized as follows. Sec.~\ref{sec:signalmodel} describes the signal model for both synchronous and asynchronous cases. Sec.~\ref{sec:CRB} investigates the position and orientation performance limits and the impact of the system geometry on localization, whereas  Sec.~\ref{sec:risphasedesign} discusses a possible design for the \ac{RIS} phase profile. In Sec.~\ref{sec:results} simulation results are reported and final conclusions are drawn in Sec.~\ref{sec:conclusions}.

Scalars (e.g., $x$) are denoted in italic, vectors (e.g., $\mathbf{x}$) in bold, and matrices (e.g., $\mathbf{X}$) in bold capital letters. $\nabla_x(a)=\partial a/\partial x$ is the partial derivative of $a$ with respect to the scalar $x$.
$\nabla_{{\mathbf x}}(\cdot)$ is the gradient operator with respect to the vector ${\mathbf x}$. Transpose and Hermitian operators are represented as $\cdot^{\mathsf{T}}$ and $\cdot^{\mathsf{H}}$, respectively. The $N \times N$ matrix with all elements being zeros  and  the $N \times N$ identity matrix are denoted by  $\mathbf{0}_{N\times N}$ and $\mathbf{I}_{N\times N}$, respectively. The operator $\operatorname{tr}\left( \mathbf{X}\right)$ denotes the trace of a matrix $\mathbf{X}$, while $\operatorname{diag}\left( \mathbf{x}\right)$ denotes a diagonal matrix with diagonal elements identified by $\mathbf{x}$. A probability density function is denoted by  $p\left(\cdot\right)$, and $\mathbb{E}\left\{ \mathbf{x} \right\}$ is the expectation of a random vector $\mathbf{x}$ with respect to its distribution. $j=\sqrt{-1}$ is the imaginary unit.

\begin{figure*}[t]
\psfrag{bo}[c][c][0.8]{$\pbo$}
\psfrag{lo}[lc][lc][0.8]{$\plo$}
\psfrag{mo}[lc][lc][0.8]{$\pmo$}
\psfrag{do}[rc][rc][0.8]{$\mathbf{p}_{\mathsf{S},0}$}	
\psfrag{x}[lc][lc][0.7]{$x$}	
\psfrag{x1}[lc][lc][0.7]{$x$}
\psfrag{y}[rc][rc][0.7]{$y$}	
\psfrag{y1}[rc][rc][0.7]{$y$}
\psfrag{z}[lc][lc][0.7]{$z$}
\psfrag{z1}[rc][rc][0.7]{$z$}
\psfrag{GG}[c][c][0.7]{\textit{Geometry of a generic station}}
\psfrag{g}[rc][rc][0.7]{$\alpha$}
\psfrag{a}[lc][lc][0.7]{$\gamma$}
\psfrag{b}[lc][lc][0.7]{$\beta$}
\psfrag{m}[lc][lc][0.8]{$\mathbf{p}_m$}
\psfrag{pb}[lc][lc][0.8]{$\mathbf{p}_b$}
\psfrag{pl}[lc][lc][0.8]{$\mathbf{p}_r$}
\psfrag{RIS}[c][c][0.9]{\textcolor{red}{$\mathbf{\acs{RIS}}$}}
\psfrag{MS}[lc][lc][0.9]{\textcolor{blue}{$\mathbf{\acs{MS}}$}}
\psfrag{B}[lc][lc][0.9]{$\mathbf{\acs{BS}}$}
\psfrag{t1}[lc][lc][0.7]{{$\thetabl$}}
\psfrag{t2}[lc][lc][0.7]{{$\thetabm$}}
\psfrag{p1}[lc][lc][0.7]{{$\philm$}}
\psfrag{p2}[lc][lc][0.7]{{$\phibm$}}
\psfrag{dBL}[rc][rc][0.7]{$\dblo$}
\psfrag{dBM}[rc][rc][0.7]{$\dbmo$}
\psfrag{dLM}[rc][rc][0.7]{$\dlmo$}
\psfrag{dlm}[lc][lc][0.7]{{$\dlm$}}
\psfrag{dlb}[rc][rc][0.7]{{$\dbl$}}
\psfrag{dbm}[c][c][0.7]{{$\dbm$}}
\psfrag{tb}[c][c][0.7]{$\theta_s$}
\psfrag{p}[c][c][0.7]{$\phi_s$}
\psfrag{db}[c][c][0.7]{$d_s$}
\psfrag{d}[c][c][0.7]{$\mathbf{p}_s$}
\centerline{
\includegraphics[width=0.6\linewidth,draft=false]{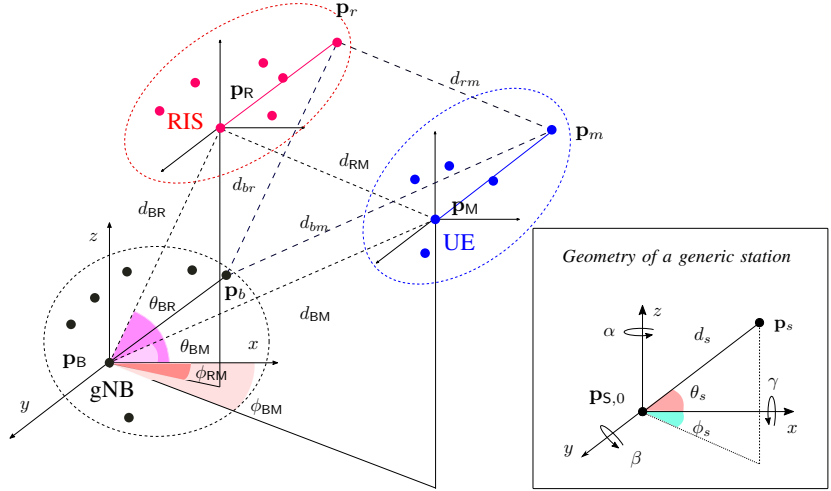} 
}
\caption{$3$D geometry of the considered localization scenario. 
}\label{fig:LISloc}
\end{figure*}
\section{System Model}
\label{sec:signalmodel}

\subsection{Localization Scenario} \label{sec:localizationscenario}
In this paper, we consider a localization scenario as in Fig.~\ref{fig:scenario} where a \ac{BS}, equipped with an antenna array with center located in position $\pbo=\left[\xbo, \ybo,\zbo \right]^{\mathsf{T}}$, performs the position and orientation estimation of a \ac{MS} with center in $\pmo=\left[\xmo, \ymo,\zmo \right]^{\mathsf{T}}$ and rotated by $\amo=\left[\alphamo, \betamo,\gammamo \right]^{\mathsf{T}}$. The geometry is reported in Fig.~\ref{fig:LISloc}. The localization is aided by the presence of a \ac{RIS}, with center located at a known position $\plo=\left[\xlo, \ylo,\zlo \right]^{\mathsf{T}}$, considered as a passive reflector that supports the \ac{BS} also for  communicating with the \ac{MS}.

%
%

According to Fig.~\ref{fig:LISloc} and considering the \ac{BS} as the center of the coordinate system, the \ac{MS} and \ac{RIS} centers' coordinates can be expressed as $\mathbf{p}_{\mathsf{S}} \triangleq \left[x_{\mathsf{S}},\, y_{\mathsf{S}},\, z_{\mathsf{S}} \right]^{\mathsf{T}}$  with $\mathsf{S} \in \left\{\mathsf{M}, \mathsf{R} \right\}$ being the label for a generic station and where the coordinates are given by
\begin{align}\label{eq:coordinates}
    &x_{\mathsf{S}}=x_{\mathsf{B}} + d_{\mathsf{BS}} \, \cos\left( \theta_{\mathsf{BS}} \right)\, \cos\left( \phi_{\mathsf{BS}} \right)\,, \\
    &y_{\mathsf{S}}=y_{\mathsf{B}} + d_{\mathsf{BS}}\, \cos\left( \theta_{\mathsf{BS}} \right)\, \sin\left( \phi_{\mathsf{BS}} \right)\,, \\
    &z_{\mathsf{S}}=z_{\mathsf{B}} + d_{\mathsf{BS}}\, \sin\left( \theta_{\mathsf{BS}}  \right)\,.
\end{align}
 Notably, the spherical coordinates can be easily retrieved from the equations above. Further, for each $\mathsf{S} \in \left\{\mathsf{B}, \mathsf{R}, \mathsf{M}\right\}$ and for each corresponding antenna index ${s} \in \left\{b,\, r,\, m \right\}$, we can indicate the antenna coordinates of each array as $\mathbf{p}_{\mathsf{S},{s}}=\mathbf{p}_{{s}}=\left[ x_s,\, y_s,\, z_s\right]^{\mathsf{T}}$ where $\forall s \in \left\{1,2, \cdots, N_{\mathsf{S}} \right\}$,
\begin{align}\label{eq:coordinates2}
    &x_{\mathsf{s}}= d_s\, \cos\left( \theta_s \right)\, \cos\left( \phi_s \right), \\
    &y_s= d_s\, \cos\left( \theta_s \right)\, \sin\left( \phi_s \right), \\
    &z_s=d_s \, \sin\left( \theta_s \right),
\end{align}
  where $N_{\mathsf{S}}$ is the number of antennas at the considered array, and $\phi_s$ and $\theta_s$ are the azimuth and elevation angles of the $s$-th antenna element measured from the array centroid, respectively.

In addition, we consider arrays that can be rotated around the axes, that is $ \forall s \in \left\{1,2, \cdots, N_{\mathsf{S}} \right\}$, we have
\begin{align}\label{eq:rotatedcoordinates}
&\mathbf{p}_{\mathsf{S},s} =\left[x_{\mathsf{S},s},\, y_{\mathsf{S},s},\, z_{\mathsf{S},s} \right]^{\mathsf{T}}= \mathbf{R}\left( \alpha_{\mathsf{S}}, \beta_{\mathsf{S}}, \gamma_{\mathsf{S}}\right)\,\mathbf{p}_{\mathsf{S},s}^{(0)}, 
\end{align}
with $\mathbf{p}_{\mathsf{S},s}^{(0)}$ being the initial array deployment, and $\mathbf{R}\left( \alpha_{\mathsf{S}}, \beta_{\mathsf{S}}, \gamma_{\mathsf{S}}\right)$ is the rotation matrix given by the multiplication of the rotation matrices for each axis and where $\left( \alpha_{\mathsf{S}}, \beta_{\mathsf{S}}, \gamma_{\mathsf{S}}\right)$ are  the roll, pitch, and yaw angles.  
The yaw is equal to the azimuth, as it is the rotation around the $z$-axis and it is here indicated with $\alpha_{\mathsf{S}}$.  $\beta_{\mathsf{S}}$ is the pitch, around the $y$-axis, whereas $\gamma_{\mathsf{S}}$ is the roll entailing a rotation around the $x$-axis. By considering counterclockwise rotations, the rotation matrix, $\mathbf{R}\left(\alpha, \beta, \gamma\right)$, is given in  \cite[(3.42)]{lavalle2006planning}.

\subsection{Signal Model for Incident Spherical Wavefronts}
We now describe a model which accounts for spherical wavefront, and it is valid also for near-field propagation conditions. In the uplink, the \ac{MS} transmits $\Ns$ \ac{OFDM} subcarriers, i.e., for the $n$-th subcarrier with  $n \in \{1,2,\cdots, \Ns\}$, we have
\begin{align}
{\mathbf x}[n]= \left[x_1,x_2, \ldots, x_{\Nm}\right]^{\mathsf{T}}\triangleq {\mathbf w}\, p[n],   
\end{align}
where $\Nm$ is the number of antennas at the \ac{MS}, $p[n]$ is the normalized data symbol corresponding to the $n$-th subcarrier with $|p|=1$, and ${\mathbf w}$ is the beamfocusing vector  given by $\mathbf {w}=[ e^{j \beta_{1}},  e^{j \beta_{2}}, \cdots,  e^{j \beta_{\Nm}}  ]^{\mathsf{T}}/\sqrt{\Nm}$
with  $\|{\mathbf w}\|=1$. Let  $\boldsymbol{\Theta} \triangleq \{ \theta_{1}, \theta_{2},\cdots, \theta_{\Nl} \}$ be the vector containing the designed phase shifts induced at the \ac{RIS}, and $\Nl$ is the number of \ac{RIS} elements. Then, we indicate with 
\begin{align}\label{eq:phaseRIS}
\bmOmega=\operatorname{diag}\left( e^{j  \boldsymbol{\Theta}}\right) \triangleq \operatorname{diag}\left( \omega_1, \ldots, \omega_r, \ldots, \omega_{\Nl}\right),
\end{align}
the $\Nl \times \Nl$ diagonal matrix containing the \ac{RIS} phases.

Differently from the \ac{RIS} literature, the \ac{BS} estimates the \ac{MS} position, $\pmo$, and its orientation, $\amo$, by exploiting also ranging and angular information present in the spherical waveform model. The received signal at the \ac{BS} for the $n$-th subcarrier can be written as 
\begin{align}\label{eq:recsign}
\y [n]&=\sqrt{P}\, \Hbm\,  \x [n]+\sqrt{P}\,  \Hbl \, \bmOmega\, \Hlm  \, \x [n]+\n [n] \nonumber\\
&\triangleq  \boldsymbol{\mu} [n]+\n [n] \,,
\end{align}
where $P$ is the signal power, $\x$ is the transmitted  vector, $\n$ is an additive thermal noise, $\Hbm$, $\Hlm$, and $\Hbl$ are the channel matrices for the \ac{BS}-\ac{MS}, \ac{RIS}-\ac{MS}, and \ac{BS}-\ac{RIS}   links, respectively.

In the following, we discriminate whether the \ac{BS} and the \ac{MS} have been synchronized or not. For non-synchronous systems, the position information can still be gathered from the spherical wavefront, even if no information can be retrieved from the \ac{TOA}.

\subsubsection{Synchronous System}

We here consider that a synchronization procedure has been performed between the \ac{BS} and the \ac{MS} prior the localization step. Once synchronized, the positioning information can be retrieved by jointly processing temporal and angular information of the received signal.
By extending \eqref{eq:recsign} to its scalar notation, the general model of the received signal can be rewritten as 
\begin{align}\label{eq:recsigsum}
 \yb[n]=&  \mu_{b}[n]  + \nb[n]  \,, &&\forall\,  b \in\{ 1,2, \cdots, \Nb\},
\end{align}
with $\Nb$ being the number of antennas at the \ac{BS} and $\nb[n]$ being the circularly symmetric zero-mean Gaussian noise with power spectral density ${\sigma}^{2}$. The useful part of the signal, without the noise, is
\begin{multline}\label{eq:mub}
    \mu_{b}[n] \triangleq \sqrt{P} \sum_{m=1}^{\Nm}  \xm[n]\, e^{-j2\pi \fn \xibm}\Big(   \rhobm  e^{-j 2\pi \fn\, \left(\tbm +\etam \right)} +\\
    +\rhoblm  \sum_{r=1}^{\Nl}  \omegal \, e^{-j2\,\pi \fn \left(\tbl+\tlm + \etal + \etam \right)}  \Big) \,,
\end{multline}
where $\fn= {n\,B}/{\Ns}$ is the considered sub-carrier, $B$ is the signal bandwidth,   and $\tbm$, $\tbl$,  $\tlm$ are the delays for each couple of antenna (e.g., $\tbm$ is the delay between the $b$th antenna at the \ac{BS} and the $m$th antenna at the \ac{MS}),  $\xibm$ is a synchronization residual (negligible for accurate synchronization procedures), and $\etam $ and $\etal$ are array non-idealities. The signal attenuation coefficients due to propagation are indicated with $\rhobm$ and $\rhoblm$ for the direct and the relayed paths, respectively.\footnote{According to the considered system geometry, the signal amplitude is about the same at each antenna as its variations are negligible.}
Since the planar wavefront approximation is not valid due to the large size of the \ac{RIS}, a spherical model is considered where the following relations hold
\begin{align}
&\tbm(\dbmo,\thetabm,\phibm)={\dbm }/{c}, \\  &\tbl(\dblo,\thetabl,\phibl)={\dbl}/{c},\\
&  \tlm(\dlmo,\thetalm,\philm)={\dlm}/{c},
\\
&\rhobm =\frac{\lambda}{4\,\pi} \, \frac{1}{\dbmo},\\
& \rhoblm =\frac{\lambda}{4\,\pi} \, \frac{1}{\dlmo+\dblo},
\end{align}
where $c$ is the speed of light, $\left(\dbmo, \thetabm, \phibm \right)$, $\left(\dblo, \thetabl,\phibl \right)$ and $\left(\dlmo, \thetalm,\philm \right)$ are the distances and angles between the \ac{BS}-\ac{MS}, \ac{BS}-\ac{RIS}, \ac{RIS}-\ac{MS} centroids, respectively, and where
\begin{align}\label{eq:dbm}
&\dbm = \sqrt{\dm^2 + \db^2 + \dbmo^2 -2\left(\gobm + \dbmo \,\gtbm\right)},
\end{align}
with $\gobm $ and $\gtbm$ containing the information of the geometry at the transmitter and at the receiver, that is
\begin{align}
\gobm& = \xb\,\xm +\yb\,\ym + \zb\,\zm \,,  \\
\gtbm &=\left(\xm-\xb \right)\cos\thetabm \cos\phibm + \nonumber \\ &+\left(\ym-\yb \right) \cos\thetabm \sin\phibm \nonumber \\
& +\left(\zm-\zb \right)\sin\thetabm.
\end{align}
The distances of arrival between the $b$th \ac{BS} antenna and the $r$th \ac{RIS} antenna and between the  $r$th \ac{RIS} antenna and the $m$th \ac{MS} antenna, namely $\dbl$ and $\dlm$, can be found using \eqref{eq:dbm} with appropriate substitutions, as done in \eqref{eq:coordinates} and \eqref{eq:coordinates2}.

Differently from traditional schemes that make the assumption of incident planar wavefront, in \eqref{eq:dbm} we  infer jointly the ranging and bearing information from the spherical waveform curvature. 
Notably, it is possible to write \eqref{eq:mub} only when the clocks of the \ac{BS}, \ac{RIS} and \ac{MS} have been synchronized. The accurate synchronization might entail several and long procedures.
In the following, we consider an asynchronous alternative where it is still possible to retrieve the \ac{MS} position from the relative phases.

\subsubsection{Asynchronous System}
As evidenced in \eqref{eq:recsigsum}, from the received signal it is possible to infer the \ac{TOA} estimate, which is possible in all those situations where a synchronization procedure has been performed.
In this case, the system is no more able to directly estimate the information of the distance from the \ac{TOA}. Instead, the incident waveform curvature, i.e.,
\begin{align}\label{eq:differential}
    &\Delta \dbm = \dbm - \dbmo=c\,\Delta\tbm, \\ 
    &\Delta \dbl = \dbl - \dblo=c\,\Delta\tbl ,
\end{align}
can be exploited for \ac{MS} localization.
%

In this case, \eqref{eq:recsign} can be written as 
%
\begin{multline}\label{eq:recsigsumasynch}
     \yb[n]=  \sqrt{P}  \,\sum_{m=1}^{\Nm}\, \xm[n]\, \Big(    \rhobm\, e^{-j\chibm}\, e^{-j 2\pi \fn\, \left( \Delta\tbm  +\etam\right)} \,  \\
     +\rhoblm\, e^{-j\chiblm}\, \sum_{r=1}^{\Nl}  \omegal \, e^{-j2\,\pi \fn \left(\Delta\tbl+\Delta\tlm + \etal + \etam \right) } \Big)+ \nb[n] ,
\end{multline}
where $\chibm$ and $\chiblm$ are uniformly distributed random variable from $0$ to $2\,\pi$  representing the phase offsets between the \ac{BS}, the \ac{MS} and the \ac{RIS} due to the lack of synchronization.


Given the proposed models for synchronous and asynchronous systems, in the following we derive the attainable fundamental performance limits.

\section{RIS-Aided Position and Orientation Error Bounds}
\label{sec:CRB}
In this section,  we derive the ultimate performance limits for the considered localization scenario. 
To this end, the \ac{CRLB} is a useful metric that represents the minimum variance of the estimation error from any unbiased estimator, and it can be represented with the inverse of the \acf{FIM} \cite{VanTrees:02,SheWin:J10}. Then, we investigate the impact of the geometry on the error through the  \ac{GDOP} metric analysis.
\subsection{The \ac{CRLB} on \ac{MS} position and orientation} \label{sec:CRLB}
Given the signal models in \eqref{eq:recsigsum} and \eqref{eq:recsigsumasynch}, we distinguish two possible estimation approaches: (i) the first one exploits a direct localization approach, and it is used for the asynchronous case; (ii) the second one is a two-stage approach that considers that the location and orientation are estimated from a set of features extracted from the signal, e.g., \acp{TOA}, \acp{AOA}, and \acp{RSSI}  \cite{Gezici:08}. 
In both cases, the parameter vector to be estimated is
\begin{equation}\label{eq:state}
\mathbf{s}=\left[\pmo,\, \amo \right]^{\mathsf{T}},
\end{equation}
where $\pmo,$ and $\amo$ contain the \ac{MS} position and orientation parameters, as indicated in \cref{sec:localizationscenario}. On the other hand, the measurement vector can be written as either  
\begin{align}
\label{eq:param1}
&{\boldsymbol \Gamma}=\mathbf{s},
\end{align}
or
\begin{align}
    &\!\!\!\! {\boldsymbol \Gamma}\!=\! \left[\rhobm, \thetabm, \phibm, \taubm,\rhoblm,  \thetalm, \philm, \taulm, \amo \right]^{\mathsf{T}},
    \label{eq:param2}
\end{align}
for  the \textit{Direct} and \textit{Two-stage} approaches, respectively,
where  $\taubm$ and $\taulm$,  $\rhobm$ and $\rhoblm$, and $\thetabm$, $\phibm$, $\thetalm$ $\philm$ are the \acp{TOA}, \ac{RSSI}, and \acp{AOA}, respectively. All the main parameters that are required to infer the location and  orientation of the \ac{MS} are included in \eqref{eq:param2}.\footnote{If the localization is only based on \ac{TOA} and \acp{AOA}, then the parameters related to \ac{RSSI} (i.e., $\rhobm$ and $\rhoblm$) can be neglected in \eqref{eq:param2}. Further, when the direct path is obstructed, all the sub-elements related to it (i.e., $\taubm$, $\phibm$, $\thetabm$, and $\rhobm$) can be discarded. Nevertheless, the resulting bound will be an upper bound on the \ac{CRLB} derived directly from the signal.} 

Note that the two approaches are the same from a \ac{CRLB} perspective. Still, we distinguish  two cases. For synchronous and asynchronous signaling, we adopt the direct and two-stage approaches, respectively.  This can be attributed to a  twofold reason: 
(i) The vector of measurements in the synchronous case allows to emphasize the parameters of the received signal which depend on the position and orientation and to quantify the error in estimating these parameters;
 (ii) 
 On the contrary, if a two-stage approach is used in the asynchronous case, where only difference of \acp{TOA} are present in \eqref{eq:differential}, then the measurement vector would consist of all the \ac{TOA} pairs between the \ac{BS}-\ac{MS} and \ac{RIS}-\ac{MS}, leading to a dimensional issue for the \ac{FIM}.
 %
 Thus, a direct approach is adopted in which the position is directly inferred from the signals received at each antenna of the \ac{BS}, allowing the measurement vector to be written in a more compact way.

Starting from \eqref{eq:param1}-\eqref{eq:param2}, the \ac{CRLB} on the \ac{MS} state vector can be written from \cite[(178)]{VanTrees:02} as
\begin{align}
\boldsymbol{\Lambda}\left( \mathbf{s} \right) \triangleq  \left[\sum_{n=1}^{\Ns} \mathbf{I}_n\left( \mathbf{s} \right) \right]^{-1},
\label{eq:crb_general}
\end{align}
where $\mathbf{I}_n\left( \mathbf{s} \right)$ is the \ac{FIM} of the state vector relative to the $n$-th subcarrier.  Hence, the 
\ac{PEB} and \ac{OEB} can be written as
\begin{align}
    &\PEB= 
    \sqrt{\operatorname{tr} \left( \left[ \boldsymbol{\Lambda}\left( \mathbf{s} \right) \right]_{1:3, 1:3} \right)},&& \OEB=
    \sqrt{\operatorname{tr} \left( \left[ \boldsymbol{\Lambda}\left( \mathbf{s} \right) \right]_{4:6, 4:6} \right)}, \label{eq:PEBOEB}
\end{align}
where $\left[ \cdot \right]_{a:b, c:d}$ indicates the sub-matrix located between rows $(a,b)$ and columns $(c,d)$. 

The \ac{FIM} can be obtained by the chain rule as \cite{Kay:93}  
\begin{equation}
\mathbf{I}_n\left( \mathbf{s}  \right) = \left(\nabla_{\mathbf{s}}  \bm{\Gamma} \right)\, \mathbf{I}_n\left( \bm{\Gamma} \right)\, \left(\nabla_{\mathbf{s}}  \bm{\Gamma} \right)^{\mathsf{T}},
\end{equation}
where $\mathbf{I}_n\left( \bm{\Gamma} \right)$ is the \ac{FIM} of the parameter vector in \eqref{eq:param1}, 
 given by \cite{VanTrees:02} as
%
\begin{equation}\label{eq:FIM_1}
\mathbf{I}_n \!\! \left(\bm{\Gamma} \right)= \mathbb{E}\left\{ \left( \nabla_{\bm{\Gamma}} \log p({\mathbf y}[n];\bm{\Gamma}) \right)^{{\mathsf{H}}} \nabla_{\bm{\Gamma}} \log p({\mathbf y}[n];\bm{\Gamma})  \right\},
\end{equation}
with $\mathbf{J} \triangleq \nabla_{\mathbf{s}}\,\,  \bm{\Gamma}$ being the Jacobian matrix  and  $\log p({\mathbf y}[n];{\boldsymbol \Gamma})$ is the log-likelihood function of the received signal vector. The log-likelihood function is computed from  \eqref{eq:recsigsum} as
\begin{align}\label{eq:likelihood}
\log p({\mathbf y}[n];{\boldsymbol \Gamma})&= -\left( {\mathbf y}[n]
-\boldsymbol{\mu}[n] \right)^{\mathsf{H}}\,\boldsymbol{\Sigma}^{-1} \, \left( {\mathbf y}[n]-\boldsymbol{\mu}[n] \right) \nonumber \\
&{\phantom{=}} -\Nb\log(\pi \sigma^2)\,,
\end{align}
where $\boldsymbol{\Sigma}=\sigma^2\, \mathbf{I}_{\Nb\times\Nb}$ is the covariance matrix of the noise. 
For the Jacobian matrix, it can be written as
\begin{align}
&{\mathbf J}=\mathbf{I}_{6 \times 6}, && \textit{Direct approach} \\
&{\mathbf J}  = \left[\begin{array}{c c} {\mathbf J}_{\pmo} & {\mathbf 0}_{3\times 3} \\ 
{\mathbf 0}_{3\times 3} & {\mathbf I}_{3\times 3} \,,
\end{array} \right], && \textit{Two-stage approach}
\end{align}
with $\mathbf{J}_{\pmo}$ indicating the term relative to the \ac{MS} position, and it is given in Appendix~\ref{app:Jacobian}.

The  elements of the \ac{FIM} in \eqref{eq:FIM_1} can be written as 
\cite{Kay:93} 
\begin{equation}\label{eq:FIMelements}
  [{\mathbf I}_n \left(\boldsymbol{\Gamma} \right)]_{i,j}=\frac{2}{\sigma^2}   \operatorname{Re}\left\{\sum_{b=1}^{\Nb} \frac{\partial \mu_{b}^{*}[n]}{\partial \Gamma_{i}} \frac{\partial \mu_{b}[n]}{\partial \Gamma_{j}}  \right\},   
\end{equation}
and their derivations are reported in Appendix~\ref{app:FIMelements}. 

Since we are considering the curvature of the wavefront in \eqref{eq:dbm}, the bound is valid also for the near-field localization that is essential when the size of arrays are sufficiently large, within the Fraunhofer distance \cite{Balanis:05}. Also, the \ac{CRLB} in \eqref{eq:crb_general} accounts for the errors due to the receiver noise (i.e., in $\sigma^2$) and the geometry of the localization scenario.

\subsection{Localization Algorithm}
One possible solution for estimating the location and orientation of the \ac{MS} is through the \ac{MLE}. 
%
The \ac{MS} location and orientation, $\mathbf{s}$, that maximize the log-likelihood function in \eqref{eq:likelihood} can be estimated as
\begin{align}
\widehat{{\mathbf s}} \triangleq \underset{\pmo \in \mathbb{R}^{3},\amo \in  {[0,2\,\pi]}^{3}}{\operatorname*{argmax}}  \log p({\mathbf y}[n];\mathbf{s}).
\end{align}
The previous optimization problem can be solved by a grid search or by iterative methods such as Newton-Raphson and expectation-maximization algorithms, however, the convergence of iterative methods to the global maximum is not guaranteed. 

The \ac{MLE} is known to approach the \ac{CRLB} derived in \cref{sec:CRLB} for asymptotically high \acp{SNR}. Also, if there exists an efficient estimator with a variance that coincides with the \ac{CRLB}, it will be the \ac{MLE}, which can be found by simultaneously solving the following equations \cite{Kay:93}
%
\begin{equation}
    {\nabla}_{\mathbf{s}}\, p({\mathbf y}[n];\mathbf{s}) = {\mathbf 0}_{6\times 1}.
\end{equation}
%
For asymptotically large number of measurements, i.e., snapshots ${\mathbf y}$  in \eqref{eq:recsign}, or high \ac{SNR}, the error in estimating the  location and orientation tends in distribution to a zero mean Gaussian distribution (indicated with ``d") with covariance matrix $\boldsymbol{\Lambda}\left( \mathbf{s} \right)$ expressed in \eqref{eq:crb_general}, i.e.,
\begin{equation}
    \widehat{{\boldsymbol{\mathsf {s}}}}-{{\mathbf s}} \overset{{\text d}}{\longrightarrow} 
    \mathcal{N}\left({\mathbf 0}_{6\times 1},\boldsymbol{\Lambda}\left( \mathbf{s} \right) \right),
\end{equation}
where $\widehat{{\boldsymbol{\mathsf {s}}}}$ is a random vector representing the estimated parameters through the \ac{MLE} \cite{Kay:93}.
%

Alternatively, the two-stage approach can be adopted, where the signal attenuation coefficients can be estimated from the \ac{RSSI}. On the other hand, the bearing angles (i.e., \acp{AOA}) can be estimated through variations of MUSIC algorithms or compressive sensing for both on- and off-grid methods with guaranteed recovery under some mild conditions. The compressive sensing based estimators have the advantage that the angles can be recovered even from a single snapshot of ${\mathbf y}$, while the performance can be further enhanced by considering multiple measurement vectors (i.e., several snapshots) \cite{YanXieZha:13,ElzanatyGioChi:19,LeeBreJun:12,ElzanatyGioChi:17}. For \ac{TOA} estimation, two possible  schemes can be considered based on correlators (i.e., matched filters) or on energy-based solutions (i.e., energy detectors) \cite{DarChoMoe:08,DarGio:09,GioChiani:13,DarCloDju:15}. The second is more practical as it can operate at sub-Nyquist rate.

\subsection{Geometry Impact on Direct RIS-aided Localization}\label{sec:GDOP}
 The \ac{CRLB} derived in Sec.~\ref{sec:CRLB} does not explicitly quantify the impact of the system geometry on the performance,
because it includes also  the effect of the input noise  \cite{yarlagadda2000gps}. Hence,  we now investigate the solely impact of the geometry on the localization performance using a \ac{GDOP} analysis. 

In particular, as a \ac{GDOP} metric, we consider the ratio between the \ac{RMSE} of position and the \ac{RMSE} of measurement (ranging) error, i.e., \cite{levanon2000lowest}
\begin{align}
    &\mathsf{GDOP}=\frac{\sqrt{\sigma_x^2+\sigma_y^2+\sigma^2_z}}{\sigma_M} \triangleq \frac{\mathsf{RMSE}\left( \mathbf{p}\right)}{\sigma_M},
\end{align}
where $\sigma_M$ is the \ac{RMSE} of the measurements, e.g., in GPS positioning it is the standard deviation of ranging measurements. Since the \ac{RMSE} is lower bounded by the \ac{CRLB}, the \ac{GDOP} can be also defined as a function of the \ac{PEB} as
\begin{align}
    &\!\sigma_M \!\cdot\! \mathsf{GDOP}= {\mathsf{RMSE}\left( \mathbf{p}\right)} \ge  {\sqrt{\operatorname{tr}\left(\mathsf{CRB}\left( \mathbf{p}\right)\right)}}=\mathsf{PEB}.
\end{align}

Differently from the parameter vector in \eqref{eq:param1}, in direct localization, position and orientation are directly estimated from the received signals at the receiver, with $\boldsymbol{\Gamma}= \mathbf{s}$ as in \eqref{eq:param2}. 
In this specific case, the measurement noise standard deviation is the same for all the antennas and corresponds to the thermal noise, i.e. $\sigma_M=\sigma$.

Given such signal-related measurements, the \ac{GDOP} can be computed from the \ac{CRLB} expression in \eqref{eq:crb_general} where the generic element in the \ac{FIM} is given by
\begin{align}\label{eq:newFIM}
    [\mathbf{I}_n\left( \mathbf{s} \right)]_{i,j}=\frac{2}{\sigma^2}   \operatorname{Re}\left\{\sum_{b=1}^{\Nb} \frac{\partial \mu_{b}^{*}[n]}{\partial \ell_{i}} \frac{\partial \mu_{b}[n]}{\partial \ell_{j}}  \right\}, 
\end{align}
where $\ell_i \in \mathbf{s}$ is either related to the position or to the orientation of the \ac{MS}. Therefore, we can write  the \ac{GDOP} for position and orientation as \cite{levanon2000lowest}
\begin{align}
\label{eq:GDOPP}
&\mathsf{GDOP}_{\pmo}\!\!=\!\!\frac{1}{\sigma\,\kappa_{\pmo}} \!\sqrt{\!\operatorname{tr}\!\left(\left[\sum_{n=1}^{\Ns} \mathbf{I}_n\left( \mathbf{s} \right) \right]^{-1}_{1:3,1:3} \right)}\!=\! \frac{\mathsf{PEB}}{\sigma\,\kappa_{\pmo}}, \\
&\mathsf{GDOP}_{\amo}\!=\!\frac{1}{\sigma\,\kappa_{\amo}} \, \sqrt{\operatorname{tr}\left(\left[\sum_{n=1}^{\Ns} \mathbf{I}_n\left( \mathbf{s} \right) \right]^{-1}_{4:6,4:6} \right)}= \frac{\mathsf{OEB}}{\sigma\,\kappa_{\amo}},
\label{eq:GDOPO}
\end{align}
where $\kappa_{\pmo} \,[\text{m}/\sqrt{\text{Watt}}]$ and $\kappa_{\amo} \,[\text{radians}/\sqrt{\text{Watt}}]$ are  the  normalization factors for the $\mathsf{GDOP}$ to become dimensionless. For example, in our settings
the  normalization factors $\kappa_{\pmo}$ and $\kappa_{\amo}$ can be designed as  $\dbmo / \sqrt{P}$ and $1/\sqrt{P}$, respectively. 
With this definition, the position and orientation errors  are proportional to the \ac{GDOP}, i.e., 
$\mathsf{PEB} \propto \sigma\, \mathsf{GDOP}_{\pmo}$ and $\mathsf{OEB} \propto \sigma\, \mathsf{GDOP}_{\amo} $\cite{levanon2000lowest}.

\section{\ac{RIS} Phase Design}
\label{sec:risphasedesign}
An important concern, when using \ac{RIS}, is the proper design of the phase profile in order to exploit as much as possible the \ac{RIS} potentialities but, unfortunately, it usually entails planar wavefronts incident to the \ac{RIS} \cite{razavizadeh20203d}.
Thus, in the following we consider possible alternatives for the design of the phase profile accounting for spherical wavefronts.


\subsubsection{Optimal RIS Phase Design}\label{sec:optimalrisphase}
The first possibility considers the optimal phase shifts induced at the \ac{RIS}  that minimize the position or orientation error bounds, i.e.,
\begin{align}\label{eq.optts2}
&\PEB^{*} \triangleq \underset{\boldsymbol{\Theta} \in \, [0,2\,\pi]^{\Nl}\,}{\text{minimize}}
 \PEB(\boldsymbol{\Theta})\,, \\
&\OEB^{*} \triangleq \underset{\boldsymbol{\Theta} \in \, [0,2\,\pi]^{\Nl}\,}{\text{minimize}}
 \OEB(\boldsymbol{\Theta}).
\end{align}
Notably, if from one side this approach is complex as it involves the minimization of the inverse of the \ac{FIM}, from the other side it represents the optimal configuration that allows to minimize the position and orientation error bounds.   


\subsubsection{Proposed RIS Phase Design}\label{sec:proposedRISphase}
Another possibility is to consider an ad-hoc approach that maximizes the sum of the \acp{SNR} at each
\ac{BS} antenna as
\begin{equation}\label{eq:optts3}
\begin{aligned}
&\underset{\boldsymbol{\Theta} \in \, [0,2\,\pi]^{\Nl}}{\text{Maximize}}
&\mathsf{SNR}(\boldsymbol{\Theta})\,,
\end{aligned}
\end{equation}
where
\begin{align}\label{eq:swinequality}
\mathsf{SNR}(\boldsymbol{\Theta})&=\frac{P}{\sigma^2}\,{\mathbf{b}}^{\mathsf{T}}{\bigg\vert    \Hbm  {\mathbf w}
+ \Hbl\bmOmega(\boldsymbol{\Theta} )\Hlm  \, {\mathbf w}  \bigg\vert}^{2}\nonumber \\
&\leq \mathsf{SNR}_\mathrm{DL}+\mathsf{SNR}_\mathrm{MP}(\boldsymbol{\Theta})\,, 
\end{align}
with
\begin{align}
  \mathsf{SNR}_\mathrm{DL}&\triangleq  \frac{P}{\sigma^2} \mathbf{b}^{\mathsf{T}}\, {\bigg\vert  \Hbm  {\mathbf w}\bigg\vert}^{2}, \nonumber\\
\mathsf{SNR}_\mathrm{MP} (\boldsymbol{\Theta})&  \triangleq \frac{P}{\sigma^2} \mathbf{b}^{\mathsf{T}}\, {\bigg\vert \Hbl \bmOmega(\boldsymbol{\Theta}   ) \Hlm \, {\mathbf w}  \bigg\vert}^{2}, 
\end{align}
\noindent where $\textbf{b}=\mathbf{1}_{\Nb\times 1}$ is a vector of all ones, $\mathsf{SNR}_\mathrm{DL}$ is the sum of the \acp{SNR} from the direct link between the \ac{BS} and the \ac{MS}, while $\mathsf{SNR}_\mathrm{MP}$ represents the sum of the \acp{SNR} from    the multipath component travelling through the \ac{RIS}. Regarding  \eqref{eq:swinequality}, it results from the Cauchy–Schwarz inequality with equality iff the phase of the direct path coincides with the phase of the reflected path. In order to design the \ac{RIS} phase profile, we operate as follows: (i)  first, we maximize  the upper bound on the \ac{SNR} in \eqref{eq:swinequality}; (ii) second, we design an additional constant phase shift for the \ac{RIS} phase profile such that the Cauchy–Schwarz inequality is satisfied with equality that is, the direct link and the multipath component are coherently summed up at each antenna. 

According to the aforementioned considerations, we  have
%
\begin{equation}
\begin{aligned}
&\underset{\boldsymbol{\Theta} \in \, [0,2\,\pi]^{\Nl}}{\text{Maximize}}
&\mathsf{SNR}_\mathrm{DL}+\mathsf{SNR}_\mathrm{MP}(\boldsymbol{\Theta})\,,
\end{aligned}
\end{equation}
that can be further simplified to
\begin{equation}\label{eq:optts5}
\begin{aligned}
&\underset{\boldsymbol{\Theta} \in \, [0,2\,\pi]^{\Nl}}{\text{Maximize}} &&\mathbf{b}^{\mathsf{T}}\left\vert \Hbl\bmOmega(\boldsymbol{\Theta} ) \Hlm \, {\mathbf w}  \right \vert^{2}\,,
\end{aligned}
\end{equation}
that gives
\begin{equation}\label{eq:optts6}
\begin{aligned}
&\underset{\boldsymbol{\Theta} \in \, [0,2\,\pi]^{\Nl}}{\text{Maximize}} &&\sum_{b=1}^{\Nb}  \left \vert \sum_{r=1}^{\Nl}  \sum_{m=1}^{\Nm}\,  e^{j \thetar} \,e^{j \betam}  e^{-j2\,\pi\, \fc \left(\tbl+\tlm\right)}  \right \vert^2,
\end{aligned}
\end{equation}
where $\fc$ is the central sub-carrier.

Unfortunately, the number of degrees of freedoms, i.e.,  the number of controllable phase shifts at the \ac{RIS}, is not enough to perfectly adjust the phase of the signals at the \ac{BS}.  
To combat such an issue, we relax the problem by minimizing the sum of square distance of the phases from their related centroid $\bar{\phi}(\boldsymbol{\Theta} ) $, inspired by the k-means algorithm \cite{Mac:67}.
Thus we write
\begin{multline}\label{eq:optts8}
\underset{\boldsymbol{\Theta} \in \, [0,2\,\pi]^{\Nl} }{\text{Minimize}} \gamma (\boldsymbol{\Theta}) \triangleq \sum_{r=1}^{\Nl} \sum_{b=1}^{\Nb} \sum_{m=1}^{\Nm}\, \,\left [ \thetar + \betam \right.  \\
\left. -2\,\pi\, \fc \left(\tbl+\tlm\right)- \bar{\phi}(\boldsymbol{\Theta} ) \right ] ^{2} \,,
\end{multline}
where $\gamma (\boldsymbol{\Theta})$ is the objective function of interest, and the centroid $\bar{\phi}(\boldsymbol{\Theta} )$ is given by
\begin{align}
&\!\!\!\bar{\phi}(\boldsymbol{\Theta}) =\frac{1}{\Nl \Nm \Nb}\! \sum_{r=1}^{\Nl}\! \sum_{b=1}^{\Nb} \!\sum_{m=1}^{\Nm}\!\left[ \thetar \!+\! \betam \!-
\!2\pi \fc \left(\tbl\!+\!\tlm\right)\right] \nonumber \\
&\!\!\!=\frac{1}{\Nl} \theta_{k}\!+\!\frac{1}{\Nl}\! \sum_{r=1, r\ne k}^{\Nl}\! \! \thetar \!+\!\frac{1}{\Nm \Nb \Nl}\! \sum_{r=1}^{\Nl}\!  \sum_{b=1}^{\Nb} \!
\sum_{m=1}^{\Nm}\! \Cbrm\,,
\end{align}
where $\Cbrm=\betam -2\,\pi \, \fc \, \left(\tbl+\tlm\right)$.
It can be verified that $\gamma (\boldsymbol{\Theta})$ is a convex function. More precisely, $\gamma (\boldsymbol{\Theta})$ is convex, as the composition of a convex function with an affine mapping is convex, and the  positive weighted sum of convex functions preserves the function convexity \cite[3.2.2]{BoydVan:04} and \cite[3.2.1]{BoydVan:04}. 

The objective function can be expressed for ${k\in 
\{1,2,\cdots, \Nl\}}$ as
\begin{align}
\gamma (\boldsymbol{\Theta}) =& \sum_{b=1}^{\Nb} \sum_{m=1}^{\Nm}\, \,{\left [ \thetak + \Cbkm - \bar{\phi}(\boldsymbol{\Theta}) \right ]}^{2} + \nonumber \\
&+ \sum_{r=1, r\ne k}^{\Nl} \sum_{b=1}^{\Nb} \sum_{m=1}^{\Nm}\, \,{\left [ \thetar + \Cbrm - \bar{\phi}(\boldsymbol{\Theta}) \right ]}^{2}.
\end{align}
Since the objective function is convex, the optimal solution can be found by solving the following equations in $\theta_k$ \cite{BoydVan:04}, 
\begin{align}
&\frac{\partial \gamma (\boldsymbol{\Theta}) }{\partial \thetak}= \sum_{b=1}^{\Nb} 
\sum_{m=1}^{\Nm}\, \ 2 \left(1-\frac{1}{\Nl}\right)\,\left [\thetak + \Cbkm- \bar{\phi}(\boldsymbol{\Theta}) \right ] + \nonumber \\
&+\sum_{r=1, r\ne k}^{\Nl} \sum_{b=1}^{\Nb} 
\sum_{m=1}^{\Nm}\,-  \frac{2}{\Nl}\,{\left [ \thetar +\Cbrm - \bar{\phi}(\boldsymbol{\Theta}) \right ]} =0,
 \end{align}
for each $k\in \{1,2,\cdots, \Nl\}$. After some manipulations, we get
\begin{align}
&\left(1-\frac{1}{\Nl}\right)\thetak-\frac{1}{\Nl} \sum_{r=1,r\ne k}^{\Nl} \thetar+\frac{1}{\Nm \Nb} \nonumber \\
& \times \left( \sum_{b=1}^{\Nb}  \sum_{m=1}^{\Nm}  \Cbkm- \frac{1}{\Nl}\,\sum_{r=1}^{\Nl} \sum_{b=1}^{\Nb}  \sum_{m=1}^{\Nm}  \Cbrm \right) =0.
\end{align}
%
Operating like this, the $\Nl$ linear equations in $\Nl$ unknowns (i.e., the \ac{RIS} phases) can be simultaneously solved to obtain the phase shifts optimized for this specific problem. An optimal, albeit not unique, phase profile that minimizes the convexified objective function in \eqref{eq:optts8}  can be written in closed-form as
\begin{align}\label{eq:optimalphases}
\hat{\theta}_{k}&= \frac{2\, \pi \, \fc}{\Nm\, \Nb}\, \sum_{b=1}^{\Nb}\sum_{m=1}^{\Nm} \left[ \tau_{bk} + \tau_{km} - \frac{1}{\Nl} \sum_{r=1}^{\Nl} \left( \tau_{br} + \tau_{rm} \right)  \right]\,,
\end{align}
where 
$\hat{\boldsymbol{\Theta}} \triangleq \{\hat{\theta}_{1},\cdots,\,\hat{\theta}_{r},\,\cdots,\,\hat{\theta}_{\Nl}  \}$.
Notably, the solution $\hat{\boldsymbol{\Theta}}$ is not unique, since adding a constant phase shift $\phi_{c}$ to the \ac{RIS} phases yields to the same value for the objective function in \eqref{eq:optts8}, as it will be accounted for also in the centroid $\bar{\phi}(\boldsymbol{\Theta} )$. Indeed, even the objective function in \eqref{eq:optts6} will not be changed by adding  a constant phase shift because of the absolute operator.

Having maximized the upper bound on the \ac{SNR}, now we derive the constant phase $\phi_{c}$ such that the Cauchy–Schwarz inequality is satisfied with equality and, hence, the \ac{SNR} is maximized. More precisely, substituting the optimal phases from \eqref{eq:optimalphases} in \eqref{eq:optts3} and maximizing it with respect to $\phi_{c}$ we get
\begin{align}\label{eq:maxsumsnr}
&\underset{\phi_{c} \in \, [0,2\,\pi] }{\text{Maximize}} \quad \frac{P}{\sigma^2}\,{\mathbf{b}}^{\mathsf{T}}{\bigg\vert \Hbm \,\, {\mathbf w} \,\,
e^{j \phi_{c}} \,\, \Hbl \,\,  \bmOmega(\hat{\boldsymbol{\Theta}}) \,\, \Hlm\,\, {\mathbf w}   \bigg\vert}^{2}\,.
\end{align}
Again, the number of degrees of freedom is not sufficient for adjusting  the phase of the direct and reflected links for all the receiving antennas at the \ac{BS}.  Hence, $\phi_{c}$  can be designed to minimize the difference between the phases of the direct and reflected link, i.e.,
%
\begin{multline}\label{eq:optphic}
\underset{{\phi_{c}} \in \, [0,2\,\pi] }{\text{Minimize}}
 \,\,\, \gamma_d(\phi_{c}) \triangleq \sum_{r=1}^{\Nl} \sum_{b=1}^{\Nb} 
\sum_{m=1}^{\Nm}\, \left [ \Big(\betam  -2\pi\, \fc\, \tbm \Big)+ \right. \\  \left. -\left(\phi_{c}+  \hat{\theta}_{r} + \betam -2\,\pi \fc \left(\tbl+\tlm\right) \right)\right ] ^{2} \,.
\end{multline}
The optimal $\phi_{c}$ can be found by solving the following equation
\begin{align}\label{eq:phic}
&\frac{\partial \gamma_d(\phi_{c})}{\partial \phi_{c}}\!=\!\sum_{r=1}^{\Nl} \sum_{b=1}^{\Nb} \sum_{m=1}^{\Nm} -2 \left [-2\pi \fc \tbm + \betam + \right.
\nonumber \\
&\left. - \left(\phi_{c}+  \hat{\theta}_{r} + \betam -2\,\pi\, \fc \left(\tbl+\tlm\right) \right)\right ]\! =0\,,
\end{align}
leading to 
\begin{align}\label{eq:hatphic}
\hat{\phi}_{c}&=
\frac{2\pi \fc }{\Nl \Nm \Nb}   \sum_{b=1}^{\Nb} \sum_{m=1}^{\Nm} \sum_{r=1}^{\Nl} \left(- \tbm + \tbl + \tlm   \right)\,,
\end{align}
obtained by substituting the derived \ac{RIS} phases in \eqref{eq:optimalphases} and distribute the summations. 

Finally, by combining \eqref{eq:optimalphases} and \eqref{eq:hatphic}, the designed phases for the \ac{RIS} that accounts for the direct and reflected paths can be written, for each $k\in \{1,2,\cdots, \Nl\}$, as
%
%
\begin{align}\label{eq:optimaltheta}
\theta^{*}_{k}&=\hat{\theta}_k+\hat{\phi}_{c} 
= \frac{2\pi \fc}{\Nm\, \Nb} \,\sum_{b=1}^{\Nb}\sum_{m=1}^{\Nm} \left( \tau_{bk} + \tau_{km} - \tbm\right).
\end{align}
%


\section{Numerical Results}\label{sec:results}

\subsection{Simulation Parameters}
According to the previous analysis, we now evaluate the attainable localization and orientation performance limits for different scenarios.
%
%
More specifically, we here focus on planar antenna array configuration,\footnote{Note that the previous analysis is valid for any geometric configuration, i.e., any antennas spatial deployment and arrays orientation.} as they allow compact deployment of massive arrays in \acp{BS} and \ac{MS}, as well as $3$D beamfocusing capabilities \cite{HonBaeKo:17,NadKamSlim:19}. Moreover, in the perspective to place \ac{RIS} on walls, planar geometry represents a practical solution \cite{di2019smart,razavizadeh20203d}. 

\begin{figure}
\psfrag{B}[rc][rc][0.7]{\ac{BS}}
\psfrag{MS}[lc][lc][0.7]{\ac{MS}}
\psfrag{RIS}[rc][rc][0.7]{\ac{RIS}}
\psfrag{XR}[lc][lc][0.6]{$\plo$}
\psfrag{XB}[lc][lc][0.6]{$\pbo$}
\psfrag{XM}[rc][rc][0.6]{$\pmo$}
\psfrag{xm}[c][c][0.5]{$x'$}
\psfrag{ym}[c][c][0.5]{$y'$}
\psfrag{zm}[rc][rc][0.5]{$z'$}
\psfrag{xb}[lc][lc][0.5]{$x''$}
\psfrag{yb}[lc][lc][0.5]{$y''$}
\psfrag{x}[c][c][0.5]{$x$}
\psfrag{y}[lc][lc][0.5]{$y$}
\psfrag{z}[rc][rc][0.5]{$z$}
\psfrag{zb}[lc][lc][0.5]{$z''$}
\psfrag{xr}[lc][lc][0.5]{$x'''$}
\psfrag{yr}[lc][lc][0.5]{$y'''$}
\psfrag{zr}[rc][rc][0.5]{$z'''$}
\psfrag{Nb}[rc][rc][0.6]{$\Nb/2$}
\psfrag{Nr}[c][c][0.6]{$\Nl/2$}
\psfrag{Nm}[rc][rc][0.6]{$\Nm/2$}
\psfrag{d}[lc][lc][0.6]{$\dant$}
\psfrag{pmi}[rc][rc][0.6]{$\mathbf{p}_{\mathsf{M},m}^{(0)}$ }
\psfrag{pri}[lc][lc][0.6]{$\mathbf{p}_{\mathsf{R},r}^{(0)}$}
\psfrag{pbi}[rc][rc][0.6]{$\mathbf{p}_{\mathsf{B},b}^{(0)}$}
\centerline{
\includegraphics[width=0.6\linewidth,draft=false]{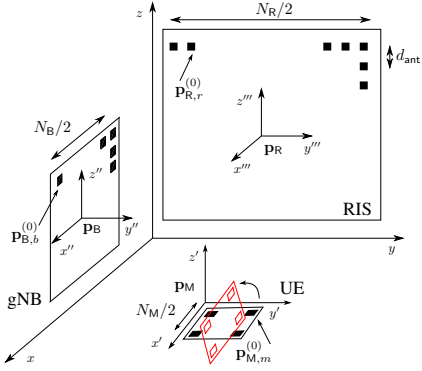}}
\caption{Considered $3$D localization scenario. An example of rotated \ac{MS} is depicted in red. 
} 
\label{fig:specialgeomtry}
\end{figure}

The  \ac{BS} is assumed to be located at the origin, i.e., at $\pbo\triangleq \left[\xbs, \ybs, \zbs\right]^{\mathsf{T}} ~\text{(m)}=\left(0,0,0\right)$, if not otherwise indicated and the initial positions of the antennas (in absence of rotations) can be  represented as in Fig.~\ref{fig:specialgeomtry} where the \ac{BS}, \ac{RIS}, and \ac{MS} are lying on the $XZ$- and $YZ$-, $XY$- planes, respectively, and the coordinates of the array elements are given by
%
\begin{figure}[t!]
\centering
\begin{subfigure}[b]{0.43\textwidth}
\centering
\psfrag{XM}[c][c][0.8]{$\xms$\, (m)}
\psfrag{YM}[c][c][0.8]{$\yms$\, (m)}
\begin{overpic}[width=0.9\linewidth,draft=false]{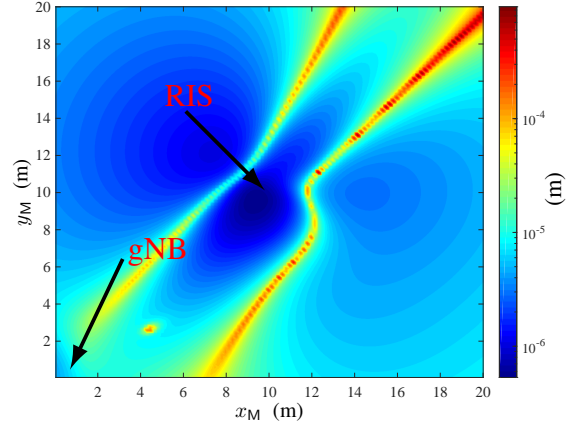} 
\put(280,590){\color{red}\large \ac{RIS}}
\put(320,580){\linethickness{0.5mm}\color{black}\vector(1,-1){150}}
\put(200,300){\linethickness{0.5mm}\color{black}\vector(-200,-420){100}}
\put(210,300){\color{red}\large \ac{BS}}
\put(1000,400){\small \rotatebox{90}{(m)}}
\end{overpic}
\caption{\ac{PEB}} \label{Fig:HeatmapPEB}
\end{subfigure}%
\vspace{0.3cm}
\qquad
\begin{subfigure}[b]{0.43\textwidth}
\centering
\psfrag{XM}[c][c][0.8]{$\xms$\, (m)}
\psfrag{YM}[c][c][0.8]{$\yms$\, (m)}
\begin{overpic}[width=0.9\linewidth,draft=false]{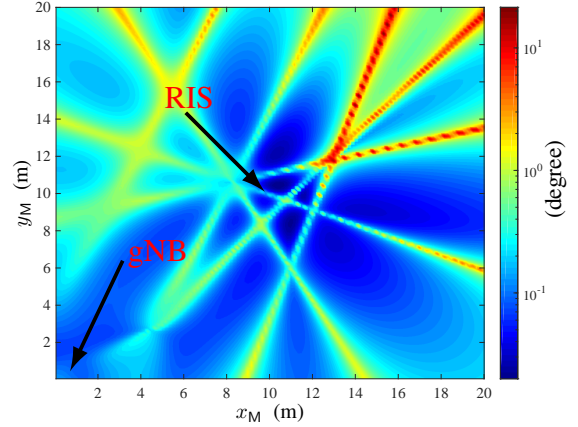} 
\put(280,590){\color{red}\large \ac{RIS}}
\put(320,580){\linethickness{0.5mm}\color{black}\vector(1,-1){150}}
\put(200,300){\linethickness{0.5mm}\color{black}\vector(-200,-420){100}}
\put(210,300){\color{red}\large \ac{BS}}
\put(1000,380){\small \rotatebox{90}{(degree)}}
\end{overpic}
\caption{\ac{OEB}} \label{Fig:HeatmapOEB}
\end{subfigure}%
\caption{(a) \ac{PEB} in meters and (b) \ac{OEB} in degrees for different mobile locations in an area of $20\times 20$ m$^2$ on the $XY$-plane. 
The orientation of the \ac{MS} is set to $\amo = \left(\pi/6, \pi/6, \pi/6 \right)$.} 
\label{Fig:Heatmap}
\end{figure}
\begin{align}
   &\mathbf{p}_{\mathsf{B},i}^{(0)}=\dant \!\left[ \bigg\lfloor \frac{i}{\sqrt{\Nb}} \bigg\rfloor\, ,0, \left( i\,  \operatorname{mod}\,  \sqrt{\Nb} \right)\, \right]^{\mathsf{T}}\!\!\!, i   \in \left\{1, \ldots, \Nb\right\}, \nonumber \\
    &\mathbf{p}_{\mathsf{R},i}^{(0)}=\dant \left[0,\, \bigg\lfloor \frac{i}{\sqrt{\Nl}} \bigg\rfloor, \left( i\,  \operatorname{mod}\,  \sqrt{\Nl} \right) \right]^{\mathsf{T}}\!\!\!, i \in \left\{1, \ldots, \Nl\right\}, \nonumber \\
    &\mathbf{p}_{\mathsf{M},i}^{(0)}\!=\!\dant \!\left[\bigg\lfloor \frac{i}{\sqrt{\Nm}} \bigg\rfloor, \left( i\,  \operatorname{mod}\,  \sqrt{\Nm} \right), 0 \right]^{\mathsf{T}}\!\!\!, i \in \left\{1, \ldots, \Nm\right\}\!,
\end{align}
%
where $\operatorname{mod}$ is the modulo operator, $\dant=\lambda/2$ is the inter-antenna spacing,
and the rotated antenna elements for a given roll, pitch and yaw angles can be defined as in \eqref{eq:rotatedcoordinates}. 
In particular, while the \ac{BS} and the \ac{RIS} are fixed on the $XZ$- and $YZ$- planes (i.e., $\alpha_{\mathsf{B}}=\beta_{\mathsf{B}}=\gamma_{\mathsf{B}}=0,\,\alpha_{\mathsf{R}}=\beta_{\mathsf{R}}=\gamma_{\mathsf{R}}=0$), respectively, the \ac{MS} can freely rotate around $x-$, $y-$, and $z-$ axis with angles $\gammamo$, $\betamo$, and $\alphamo$, respectively, according to Fig.~\ref{fig:LISloc}.

At the transmitter, we considered \ac{OFDM} signaling and for each simulation scenario we specify the number $\Ns$ of adopted subcarriers, with unity transmitted power $P=1\,$Watt,  a  carrier frequency $\fc=28\,$GHz and a signal bandwidth {$B= \Ns\, \Delta f$}, where {$\Delta f=240$~kHz} being the subcarrier spacing \cite{ZaiAthChe:18}.  
At the receiver, we set a noise figure $\mathsf{F}=3\,$dB. 

Concerning the \ac{RIS} phase profile in \eqref{eq:phaseRIS}, in the next, we use the following \textit{labels} according to the type of design: (i) \textit{Mirror}, when the \ac{RIS} does not induce any phase shift, that is $\boldsymbol{\Theta}=\mathbf{0}_{1\times \Nl}$; (ii) \textit{Random}, when the \ac{RIS} phase shifts are uniformly distributed between $0$ and $2\,\pi$; (iii) \textit{Proposed}, according to the analysis of Sec.~\ref{sec:proposedRISphase} maximizing the \ac{SNR}; (iv) \textit{Optimized \ac{CRLB}}, according to the minimization of the \ac{CRLB} reported in Sec.~\ref{sec:optimalrisphase}, and; (v) \textit{Quantized}, that accounts for $4$ quantization levels in the representation of the optimized \ac{CRLB}.

\subsection{Numerical Results}
\paragraph{PEB and OEB for Different Mobile Positions}
Fig.~\ref{Fig:Heatmap} shows the position and orientation errors in \ac{RIS}-assisted architecture by varying the \ac{MS} location in different points of the area. In particular, the location of the \ac{RIS} is  $\plo \triangleq \left[\xls, \yls,\zls \right]^{\mathsf{T}}~\text{(m)} =\left(10, 10,-1 \right)$, whereas the \ac{BS} is placed in $\pbo=\left(0,0,0\right)$, unless stated otherwise. The \ac{BS}, \ac{RIS}, and \ac{MS} are equipped with planar antenna arrays with $\Nb=36$, $\Nl=100$, $\Nm=4$ antennas, respectively. 
The \ac{MS} altitude is set to $\zms=-3\,$m, the \ac{MS} orientation to $\amo \triangleq \left[\alphamo, \betamo, \gammamo\right]^{\mathsf{T}}~\text{(rad)}=\left(\pi/6, \pi/6, \pi/6\right)$, $\Ns= 1$, and the \textit{proposed} phase design is adopted  for the \ac{RIS} phases.
%
%
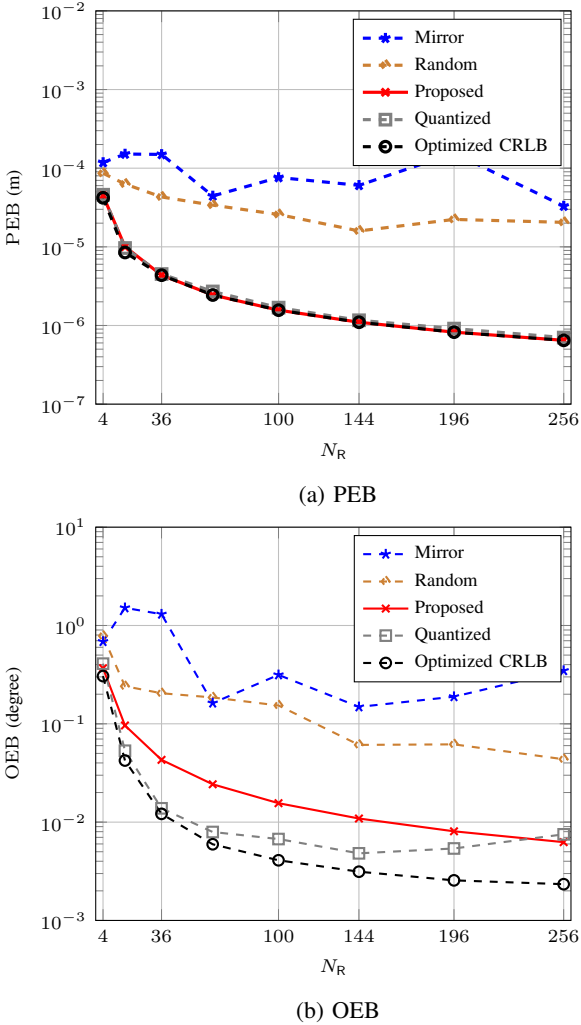
\begin{figure}[t]
	\centering
	\begin{subfigure}[b]{0.44\textwidth}
		\centering
		\pgfplotsset{every axis/.append style={
		legend style={at={(0.98,0.98)}},legend cell align=left,font=\scriptsize},
} 
\begin{tikzpicture}[trim axis left, trim axis right]
\begin{semilogyaxis}[
xlabel near ticks,
ylabel near ticks,
grid=major,
xlabel={$\Nl$},
ylabel={$\mathrm{PEB}$ (m)},
width=0.99\linewidth,
legend entries={Mirror,
	Random,
	Proposed,
	Quantized,
	Optimized \ac{CRLB}
},
	xmin= 0, xmax=260,
xtick={4,36,100,144,196, 256},
	ymin=1e-7, ymax=1e-2,
]
\addplot[blue,dashed,mark=star,very thick] table {Figures/PhaseDesignFig/PEB_Mirror.csv};
\addplot[orange!20!brown,mark=diamond,dashed,very thick] table {Figures/PhaseDesignFig/PEB_Random.csv};
\addplot[red,solid,mark=x,very thick] table {Figures/PhaseDesignFig/PEB_Proposed.csv};
\addplot[gray,dashed,mark=square,mark options=solid,very thick] table {Figures/PhaseDesignFig/PEB_QuantizedOpt.csv};
\addplot[black,mark=o,dashed,mark options=solid,very thick] table {Figures/PhaseDesignFig/PEB_Opt.csv};
\end{semilogyaxis}
\end{tikzpicture}
		\caption{\ac{PEB}} \label{Fig:PhaseDesignFigPEB}
	\end{subfigure}
	~
	\begin{subfigure}[b]{0.44\textwidth}
		\centering
		\pgfplotsset{every axis/.append style={
		legend style={at={(0.98,0.98)}},legend cell align=left,font=\scriptsize},
} %
\begin{tikzpicture}[trim axis left, trim axis right]
\begin{semilogyaxis}[
xlabel near ticks,
ylabel near ticks,
grid=major,
xlabel={$\Nl$},
ylabel={$\mathrm{OEB}$ (degree)},
width=0.99\linewidth,
legend entries={Mirror,
	Random,
	Proposed ,
	Quantized,
	Optimized \ac{CRLB}
},
%
	xmin= 0, xmax=260,
	xtick={4,36,100,144,196, 256},
ymin=1e-3, ymax=1e1,
]
\addplot[blue,dashed,mark=star,thick] table {Figures/PhaseDesignFig/OEB_Mirror.csv};
\addplot[orange!20!brown,mark=diamond,dashed,thick] table {Figures/PhaseDesignFig/OEB_Random.csv};
\addplot[red,solid,mark=x,thick] table {Figures/PhaseDesignFig/OEB_Proposed.csv};
\addplot[gray,mark=square,dashed,mark options=solid,thick] table {Figures/PhaseDesignFig/OEB_QuantizedOpt.csv};
\addplot[black,mark=o,dashed,mark options=solid,thick] table {Figures/PhaseDesignFig/OEB_Opt.csv};
\end{semilogyaxis}
\end{tikzpicture}
		\caption{\ac{OEB}}\label{Fig:PhaseDesignFigOEB}
	\end{subfigure}
	\caption{PEB and OEB vs. number of \ac{RIS} elements, namely $\Nl$, for different \ac{RIS} phase design strategies.}
	\label{Fig:PhaseDesignFig}
\end{figure}
As it can be seen in \cref{Fig:HeatmapPEB} and \cref{Fig:HeatmapOEB}, the \ac{PEB} and the \ac{OEB} are lower in proximity of the \ac{BS} and of the \ac{RIS}, with an error of about $6 \times 10^{-6}\,$~m for the position and of $0.2^{\circ}$ for the orientation when the \ac{MS} is placed at $\pmo \triangleq \left[\xms, \yms, \zms \right]^{\mathsf{T}}~\text{(m)} =(4,4,-3)$. Notably, the achieved errors depend not only on the distance from the \ac{BS} and from the \ac{RIS}, but also on the relative \ac{MS} location with respect to them, e.g., the \ac{MS} location has an effect on the actual bearing angles $\philm$ and $\phibm$ and, in turns, on the localization. 
\paragraph{PEB and OEB for Different \ac{RIS} Configurations}
\cref{Fig:PhaseDesignFig} reports the localization and orientation errors for different  number of antennas at the \ac{RIS} and different phase design strategies. We set the number of \ac{BS} antennas to $\Nb= 16$, the number of \ac{MS} antennas to $\Nm= 4$, the \ac{RIS} and \ac{MS} centroids to  $\plo=\left(4,3,1\right)$ and $\pmo=\left(5, 2,-1\right)$, respectively. The \ac{MS}  orientation around the $z$-axis is $\alphamo= \pi/6$ and a single sub-carrier is used, i.e. $ \Ns= 1$.
As previously discussed, the \textit{Optimized CRLB} phase design strategy is obtained by numerically minimizing the \ac{PEB} and \ac{OEB}. Because the \ac{CRLB} optimization problem is non-convex, the algorithm could converge to a local minimum if the initial point is far from the true solution. Therefore, we included the proposed closed-form phase design in \eqref{eq:optimaltheta} as a possible initialization for the optimization algorithm in the \textit{Optimized CRLB} phase design. 
For the \ac{PEB} in \cref{Fig:PhaseDesignFigPEB}, we can see that the \textit{proposed} design almost coincides with the \textit{Optimized \ac{CRLB}}, and that the quantization does not significantly decrease the performance. Regarding the \ac{OEB} in \cref{Fig:PhaseDesignFigOEB}, the \textit{optimized \ac{CRLB}} and its \textit{quantized} version allow to outperform the \textit{proposed} scheme. 
Another interesting aspect is that the error tends to slowly decrease for $\Nl\ge 100$, thus permitting to relax the number of antennas at the \ac{RIS} side while obtaining the good localization performance.


%

%
\begin{figure}[t]
\centering
\input{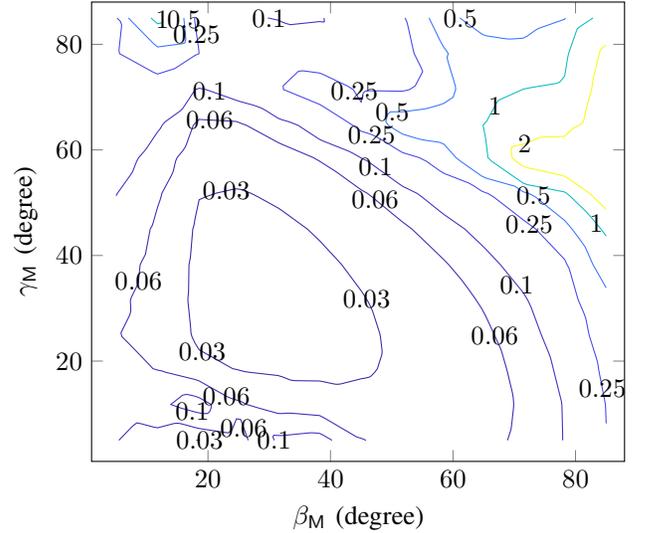}
\caption{OEB contours as a function of the orientation angles $\betamo$ and $\gammamo$ in degrees for a fixed $\alphamo=30^\circ$.}
\label{Fig:ContoursOEB}
\end{figure}
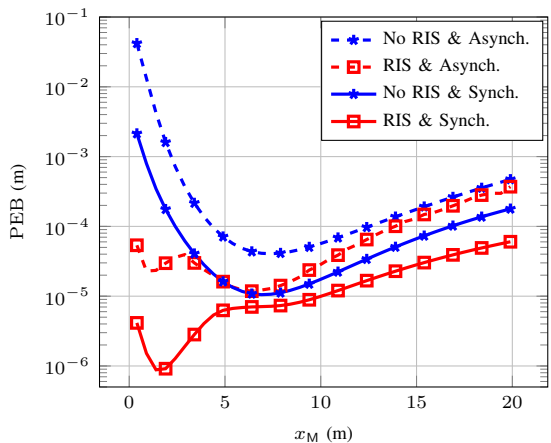
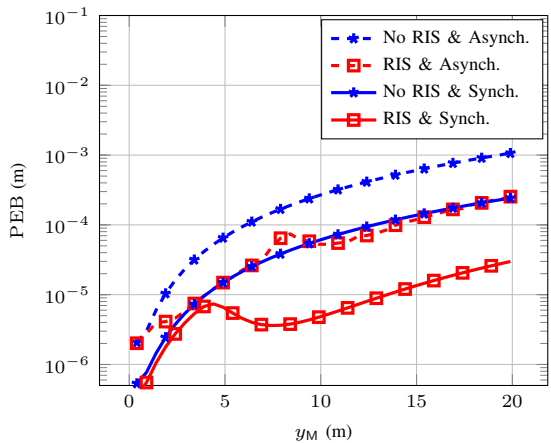
\begin{figure}[t!]
\centering
\begin{subfigure}[b]{0.42\textwidth}
\pgfplotsset{every axis/.append style={
		legend style={at={(0.99,0.99)}},legend cell align=left,font=\scriptsize}
} %
\begin{tikzpicture}
\begin{semilogyaxis}[
xlabel near ticks,
ylabel near ticks,
grid=major,
xlabel={$\xms$ (m)},
ylabel={$\mathrm{PEB}$ (m)},
width=0.99\linewidth,
legend entries={No \ac{RIS} \& Asynch.,
\ac{RIS} \&	Asynch.,
	No RIS \& Synch.,
\ac{RIS} \& Synch.,
},
ymin=5e-7, ymax=1e-1,
]
\addplot[blue,mark=star,very thick,mark options=solid,mark repeat=3,dashed] table {Figures/AsynchFigAnna/NoRISAsynchPEB_Proposed.csv};
\addplot[red,dashed,mark=square,very thick,mark options=solid,mark repeat=3,dashed] table {Figures/AsynchFigAnna/AsynchPEB_Proposed.csv};
\addplot[blue,mark=star,very thick,mark options=solid,mark repeat=3] table {Figures/AsynchFigAnna/NoRISPEB_Proposed.csv};
\addplot[red,mark=square,very thick,mark options=solid,mark repeat=3] table {Figures/AsynchFigAnna/PEB_Proposed.csv};



\end{semilogyaxis}
\end{tikzpicture}
\caption{$\pmo=\left(\xms,5,1\right)$}
\end{subfigure}
\begin{subfigure}[b]{0.42\textwidth}
\pgfplotsset{every axis/.append style={
		legend style={at={(0.99,0.99)}},legend cell align=left,font=\scriptsize}
} %
\begin{tikzpicture}
\begin{semilogyaxis}[
xlabel near ticks,
ylabel near ticks,
grid=major,
xlabel={$\yms$ (m)},
ylabel={$\mathrm{PEB}$ (m)},
width=0.99 \linewidth,
legend entries={No \ac{RIS} \& Asynch.,
	\ac{RIS} \&	Asynch.,
	No RIS \& Synch.,
	\ac{RIS} \& Synch.,
},
ymin=5e-7, ymax=1e-1,
]
\addplot[blue,mark=star,very thick,mark options=solid,mark repeat=3,dashed] table {Figures/AsynchFigAnna/YNoRISAsynchPEB_Proposed.csv};
\addplot[red,dashed,mark=square,very thick,mark options=solid,mark repeat=3,dashed] table {Figures/AsynchFigAnna/YAsynchPEB_Proposed.csv};
\addplot[blue,mark=star,very thick,mark options=solid,mark repeat=3] table {Figures/AsynchFigAnna/YNoRISPEB_Proposed.csv};
\addplot[red,mark=square,very thick,mark options=solid,mark repeat=3] table {Figures/AsynchFigAnna/YPEB_Proposed.csv};



\end{semilogyaxis}
\end{tikzpicture}
\caption{$\pmo=\left(5,\yms,1 \right)$}
\end{subfigure}
\caption{\ac{PEB} for synchronous and asynchronous signal models, for 
\ac{MS} orientation fixed to $\amo=\left( \pi/4, \pi/2, 0 \right)$. 
}
\label{Fig:Synch1}
\end{figure}

\begin{figure}[t!]
\centering
\begin{subfigure}[b]{0.42\textwidth}
\pgfplotsset{every axis/.append style={
		legend style={at={(0.99,0.99)}},legend cell align=left,font=\scriptsize}
} %
\begin{tikzpicture}
\begin{semilogyaxis}[
xlabel near ticks,
ylabel near ticks,
grid=major,
xlabel={$\xms$ (m)},
ylabel={$\mathrm{PEB}$ (m)},
width=0.99\linewidth,
legend entries={No \ac{RIS} \& Asynch.,
\ac{RIS} \&	Asynch.,
	No RIS \& Synch.,
\ac{RIS} \& Synch.,
},
ymin=7e-7, ymax=1e0,
]
\addplot[blue,mark=star,very thick,mark options=solid,mark repeat=3] table {Figures/AsynchFigGuidi/NoRISAsynchPEB_Proposed.csv};
\addplot[red,dashed,mark=square,very thick,mark options=solid,mark repeat=3] table {Figures/AsynchFigGuidi/AsynchPEB_Proposed.csv};
\addplot[blue,dashed,mark=star,very thick,mark options=solid,mark repeat=3] table {Figures/AsynchFigGuidi/NoRISPEB_Proposed.csv};
\addplot[red,mark=square,very thick,mark options=solid,mark repeat=3] table {Figures/AsynchFigGuidi/PEB_Proposed.csv};



\end{semilogyaxis}
\end{tikzpicture}
\caption{$\pmo=\left(\xms,5,1 \right)$}
\label{Fig:AvgSSynchX}
\end{subfigure}
\begin{subfigure}[b]{0.42\textwidth}
\pgfplotsset{every axis/.append style={
		legend style={at={(0.99,0.99)}},legend cell align=left,font=\scriptsize}
} %
\begin{tikzpicture}
\begin{semilogyaxis}[
xlabel near ticks,
ylabel near ticks,
grid=major,
xlabel={$\yms$ (m)},
ylabel={$\mathrm{PEB}$ (m)},
width=0.99 \linewidth,
legend entries={No \ac{RIS} \& Asynch.,
	\ac{RIS} \&	Asynch.,
	No RIS \& Synch.,
	\ac{RIS} \& Synch.,
},
ymin=7e-7, ymax=1e0,
]
\addplot[blue,mark=star,very thick,mark options=solid,mark repeat=3] table {Figures/AsynchFigGuidi/YNoRISAsynchPEB_Proposed.csv};
\addplot[red,dashed,mark=square,very thick,mark options=solid,mark repeat=3] table {Figures/AsynchFigGuidi/YAsynchPEB_Proposed.csv};
\addplot[blue,dashed,mark=star,very thick,mark options=solid,mark repeat=3] table {Figures/AsynchFigGuidi/YNoRISPEB_Proposed.csv};
\addplot[red,mark=square,very thick,mark options=solid,mark repeat=3] table {Figures/AsynchFigGuidi/YPEB_Proposed.csv};



\end{semilogyaxis}
\end{tikzpicture}
\caption{$\pmo=\left(5,\yms,1 \right)$}
\end{subfigure}
\caption{\ac{PEB} for synchronous and asynchronous signal models, averaged for 
different \ac{MS} orientation.
}
\label{Fig:Synch}
\end{figure}
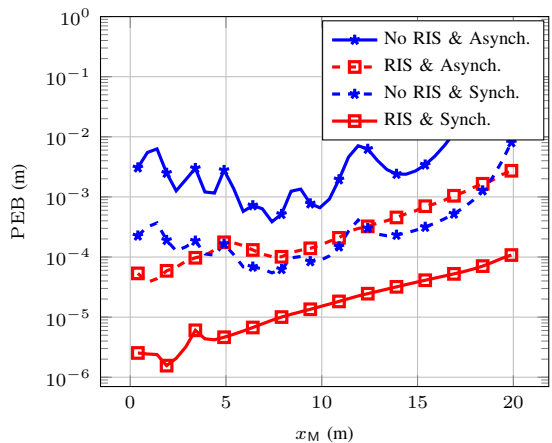
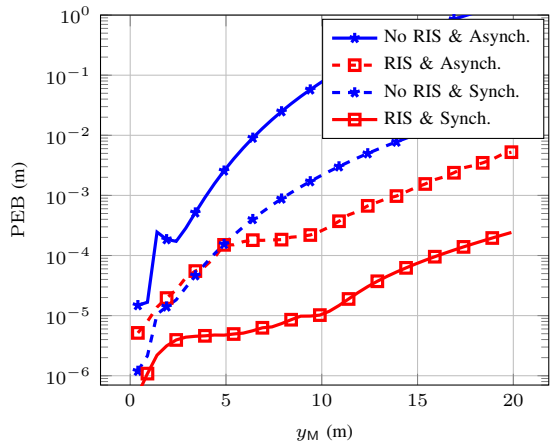

\paragraph{Analysis of the \ac{MS} Orientation}
We now analyze the impact of the mobile orientation angle on the \ac{OEB} when the location of the mobile and its orientation with respect to the $z$-axis are fixed, i.e., $\pmo=\left(15,5,-3\right)$ and $\alphamo=\pi/6$, respectively, while the orientation of the \ac{MS} around both $x$- and $y$- axis (i.e., $\betamo$ and $\gammamo$) are varied from $0$ to $\pi/2$. The number of antennas at the \ac{BS}, \ac{MS} and \ac{RIS} are $\Nb=36$, $\Nm=4$, and $\Nl=256$, respectively. The \ac{RIS} position is $\plo=\left(10, 10,-1\right)$, and the number of sub-carriers is $\Ns=8$. In this sense, according to the results reported in \cref{Fig:ContoursOEB}, we can observe that the \ac{OEB} increases when the mobile is parallel and/or perpendicular to the \ac{RIS} and the \ac{BS}. On the other hand, the \ac{OEB} decreases when $\gammamo$ and $\betamo$ are close to $30^\circ$.

%

\paragraph{Synchronous vs. Asynchronous Signaling}
We now compare the achievable performance of synchronous and asynchronous systems in an environment with and without \ac{RIS}. More specifically, we evaluated the \ac{PEB} as a function of the $x/y$-coordinates of the mobile for a fixed \ac{MS} orientation, i.e., $\amo=\left( \pi/4, \pi/2, 0 \right)$ (see \cref{Fig:Synch1}), and for averaged orientations, i.e., for $36$ configurations where both $\alphamo$ and $\betamo$ are varied between $0^\circ$ and $90^\circ$ degrees with a step of $15^\circ$ degree (see \cref{Fig:Synch}). In both configurations, the \ac{BS} and the \ac{RIS} are located at $\pbo=\left(5,0, 1.5 \right)$ and $\plo=\left(0, 5, 2 \right)$, respectively. The number of antennas and of sub-carriers are $\Nb=36$, $\Nl=64$, $\Nm=4 $, and $\Ns=8$, respectively.
In particular, \cref{Fig:Synch1}-(a)  and \cref{Fig:Synch}-(a) are obtained by fixing $\yms=5$~m, with $\xms$ spanning from $0$ to $20$~m, whereas in \cref{Fig:Synch1}-(b) and in \cref{Fig:Synch}-(b) we set $\xms=5$~m and $\yms$ is changed from $0$ to $20$~m. We can see that the \ac{PEB} decreases in the area between the \ac{BS} and the \ac{RIS} for the synchronous case, whereas there is not a significant variation for the asynchronous one. 
Also, it can be noticed that the \ac{RIS} improves the performance of the localization with up to one- and two- orders of magnitude for the synchronous and asynchronous signaling, respectively.  When the orientation is fixed, it is evident that the location of the minimum error coincides with the \ac{RIS} position for the synchronous case, i.e., $\xls=5$~m.  

The orientation of the mobile impacts the localization performance, as it can be seen in \cref{Fig:Synch}. Hence, we averaged over several mobile orientations to study the effect of the distance on the localization, regardless of the mobile orientation. As expected, when increasing the \ac{MS} distance from the \ac{BS}, e.g., by varying the $y$-coordinate,  the localization error increases faster to what happens by moving along the $x$-axis (i.e., far from the \ac{RIS}). 
\begin{figure}
    \centering
    \pgfplotsset{every axis/.append style={
		legend style={at={(0.98,0.98)}},legend cell align=left,font=\footnotesize},
} %
\begin{tikzpicture}[trim axis right,trim axis left]
\begin{semilogyaxis}[
xlabel near ticks,
ylabel near ticks,
grid=major,
xlabel={$\xms$ (m)},
ylabel={$\PEB$ (m)},
width=\figurewifthlatex\linewidth,
legend style={cells={align=left}}, 
legend entries={
\acp{RSSI} \& \acp{AOA},
\acp{TOA} \& \acp{AOA},
All parameters \\
(\acp{RSSI} \& \acp{TOA} \& \acp{AOA}),
},
	ymin=1e-5, ymax=1e-1,
]
\addplot[blue,mark=star,very thick,mark repeat=2] table {Figures/CRLBSubParameters/PEB_NoToA.csv};
\addplot[brown!60!black,mark=diamond,dashed,very thick,mark size=3,mark options=solid,mark repeat=2] table {Figures/CRLBSubParameters/PEB_NoRSSI.csv};
\addplot[red,solid,mark=x,very thick,mark repeat=2] table {Figures/CRLBSubParameters/PEB_All.csv};
%
%

\end{semilogyaxis}
\end{tikzpicture}
    \caption{\ac{PEB} for different $x$-coordinates of a \ac{MS} located in $\pmo=\left(\xms, 2, -3\right)$. The \ac{RIS} is located at $\plo=\left(4, 4, -1\right)$.
    } 
    \label{Fig:PEBSubparameters}
\end{figure}
\paragraph{Two-stage Localization}
In \cref{Fig:PEBSubparameters}, the localization accuracy is investigated for the case that the system can estimate only a subset of the parameters in \eqref{eq:param2}.

We differentiate between two cases: i) the \ac{RSSI} and \ac{AOA} are estimated; ii) the \acp{TOA} and the \ac{AOA} are estimated. Then, the two scenarios are compared with the benchmark, where the system can estimate all the parameters in \eqref{eq:param2}, and the corresponding \ac{PEB} is calculated as in \eqref{eq:PEBOEB}. The \ac{PEB} for the three cases is depicted in \cref{Fig:PEBSubparameters} for various values of mobile locations, along the $x$-axis. In the considered scenario, we set $\pmo=\left(\xms,2,-3\right)$, $\plo= \left(4, 4, -1 \right)$, $\Nb=16$, $\Nl=36$, $\Nm=4 $, and $\Ns=8$. We can see that discarding the \acp{RSSI} from the parameter vector (i.e., not relying on measuring the \acp{RSSI} for positioning purposes) has negligible impact on the \ac{PEB}. On the contrary, if the system is able only to estimate the \ac{RSSI} and not the \acp{TOA},  the localization error increases up to two order of magnitude. Therefore, in our considered setting, localization systems with accurate \ac{TOA} estimation can achieve higher performance compared to those relying on \acp{RSSI}. 
\begin{figure}[t!]
\centering
\begin{subfigure}[b]{0.4\textwidth}
\centering
\begin{tikzpicture}[trim axis left, trim axis right]
\begin{semilogyaxis}[
grid=major,
xlabel={$\xms$ (m)},
ylabel={$\sigma_{\Gamma_{j}}$ (sec)},
width=0.99\linewidth,
legend entries={
$\taubm$,  
$\taulm$
},
	xmin= 0, xmax=10,
]
\addplot[blue,mark=star,very thick,mark repeat=2] table {Figures/FIMParametersAnna/VarParm_taubm.csv};
\addplot[brown!60!black,mark=diamond,dashed,very thick,mark size=3,mark options=solid,mark repeat=2] table {Figures/FIMParametersAnna/VarParm_taulm.csv};
\end{semilogyaxis}
\end{tikzpicture}
\caption{Error bounds on the \ac{TOA} estimation}
\vspace{0.3cm}
\label{Fig:ParametersEBtime}
\end{subfigure}
\quad
\begin{subfigure}[b]{0.4\textwidth}
\centering

\begin{tikzpicture}[trim axis left, trim axis right]
\begin{semilogyaxis}[
grid=major,
xlabel={$\xms$ (m)},
ylabel={$\sigma_{\Gamma_{j}}$ (radian)},
width=0.99 \linewidth,
legend entries={
$\thetabm$, 
$\phibm$,
$\thetalm$,
$\philm$
},
	xmin= 0, xmax=10,
]
\addplot[blue,mark=star,very thick,mark repeat=2] table {Figures/FIMParametersAnna/VarParm_thetabm.csv};
\addplot[brown!40!black,mark=diamond,dashed,very thick,mark size=3,mark options=solid,mark repeat=2] table {Figures/FIMParametersAnna/VarParm_phibm.csv};
\addplot[red,mark=star,very thick,mark repeat=2] table {Figures/FIMParametersAnna/VarParm_thetalam.csv};
\addplot[green!60!black,mark=diamond,dashed,very thick,mark size=3,mark options=solid,mark repeat=2] table {Figures/FIMParametersAnna/VarParm_philm.csv};
\end{semilogyaxis}
\end{tikzpicture}
\caption{Error bounds on the angles estimation }
\label{Fig:ParametersEBAnlges}
\end{subfigure}
\caption{The error in estimating the quantities in the parameters vector vs the $x$-coordinates of the mobile location, for  $\pbo=\left(5, 0, 1.5 \right)$, $\plo=\left( 0, 5, 2 \right)$, $\amo=\left(\pi/4, \pi/2, 0 \right)$, $\Nb=36$, $\Nl=64$, $\Nm=4$, and $\Ns=8$.}
\label{Fig:ParametersEB}
\end{figure}

For the two-stage approach in estimating the location, it is beneficial to quantify the minimum possible error for estimating the parameters in \eqref{eq:param2}. To this purpose, the error bound on the parameters can be written as 
\begin{align}\label{eq:crb_parameters}
&{\sigma}_{\Gamma_{j}} \triangleq  \sqrt{ \left[\sum_{n=1}^{\Ns} \mathbf{I}_n\left( \boldsymbol{\Gamma} \right) \right]^{-1}_{j,j}}, && \forall j\in \{1,2,\cdots, |\boldsymbol{\Gamma}| \},
\end{align}
where $\mathbf{I}_n\left( \boldsymbol{\Gamma} \right)$ is the \ac{FIM} of the parameters for a given subcarrier $n$ as expressed in \eqref{eq:FIM_1} and $|\boldsymbol{\Gamma}|$ is the number of parameters in $\boldsymbol{\Gamma}$. In this regard, we depict in \cref{Fig:ParametersEB} the error for estimating the parameters  $\thetabm, \phibm, \taubm, \thetalm, \philm, \taulm$. In particular, the errors in estimating the time, i.e., ${\sigma}_{\taubm}$ and ${\sigma}_{\taulm}$, are shown in \cref{Fig:ParametersEBtime}, while the standard deviation of  the estimation errors of the angles, i.e., ${\sigma}_{\thetabm}$, ${\sigma}_{\phibm}$ and ${\sigma}_{\thetalm}$, ${\sigma}_{\philm}$, are presented in \cref{Fig:ParametersEBAnlges}. We can see that the parameters that depend on the \ac{BS}, i.e.,  ${\sigma}_{\taubm}, {\sigma}_{\thetabm}$, ${\sigma}_{\phibm}$, has minimum estimation errors near the \ac{BS} location. On the other hand, the parameters were the \ac{RIS} is involved increases as \ac{MS} gets far from the \ac{RIS}, then decreases again as the it approaches the \ac{BS}.

\begin{figure}[t!]
\centering
\begin{subfigure}[b]{0.45\textwidth}
\centering
\pgfplotsset{every axis/.append style={
		legend style={at={(0.95,0.95)}},legend cell align=left,font=\footnotesize},
} %
\begin{tikzpicture}
\begin{axis}[
xlabel near ticks,
ylabel near ticks,
grid=major,
ylabel={$\mathsf{GDOP}_{\pmo}$},
xlabel={$\Nl$},
width=0.99\linewidth,
legend entries={
$\phibm=1^\circ$,
$\phibm=45^\circ$,
$\phibm=89^\circ$,
},
]
\addplot[blue,dashed,mark=star,very thick] table {Figures/GDOPFig/GDOPPEPphi1.dat};
\addplot[orange!20!brown,mark=diamond,dashed,very thick] table {Figures/GDOPFig/GDOPPEPphi45.dat};
\addplot[red,solid,mark=x,very thick] table {Figures/GDOPFig/GDOPPEPphi89.dat};


\end{axis}
\end{tikzpicture}
\caption{The \ac{GDOP} for the mobile position.}
\vspace{0.3cm}
\label{Fig:GDOPPEB}
\end{subfigure}
\begin{subfigure}[b]{0.45\textwidth}
\centering
\pgfplotsset{every axis/.append style={
		legend style={at={(0.95,0.95)}},legend cell align=left,font=\footnotesize},
} %
\begin{tikzpicture}
\begin{axis}[
grid=major,
xlabel={$\Nl$},
ylabel={$\mathsf{GDOP}_{\amo}$},
width=0.99\linewidth,
legend entries={
	$\phibm=1^\circ$,
	$\phibm=45^\circ$,
	$\phibm=89^\circ$,
	},
	]
	\addplot[blue,dashed,mark=star,very thick] table {Figures/GDOPFig/GDOPOEPphi1.dat};
	\addplot[orange!20!brown,mark=diamond,dashed,very thick] table {Figures/GDOPFig/GDOPOEPphi45.dat};
	\addplot[red,solid,mark=x,very thick] table {Figures/GDOPFig/GDOPOEPphi89.dat};
	
	
	\end{axis}
	\end{tikzpicture}
\caption{The \ac{GDOP} for the orientation angles.}
\label{Fig:GDOPOEB}
\end{subfigure}
\caption{The \ac{GDOP} for various $\Nl$ and $\phibm$, for $\Nb= 36$, $\Nm= 16$, $\plo=\left(4, 3,1 \right)$, $\dbmo\!=\! 3~\text{m}, \alphamo= 30$, and {$\Ns\!=\! 2400$}.}
	\label{Fig:GDOP}
\end{figure}
\paragraph{Geometric dilution of precision} In \cref{Fig:GDOP}, the impact of the geometry on the localization and orientation estimation errors is investigated as a function of the number of elements in the \ac{RIS}. The \ac{GDOP} value can be considered as an amplification of the estimation error due to the geometry. Therefore, smaller values of the \ac{GDOP} indicate a favorable geometry of the mobile with respect to both the \ac{BS} and the \ac{RIS}. The \ac{GDOP} for the position, $\mathsf{GDOP}_{\pmo}$, is depicted in \cref{Fig:GDOPPEB} as a function of the number of \ac{RIS} elements for various \ac{MS} locations with different azimuth angles between the \ac{BS} and \ac{MS}, $\phibm$, while the corresponding distance and the elevation angle are fixed to $\dbmo=3$ m and $\thetabm=30^{\circ}$, respectively. It can be noticed that increasing the number of \ac{RIS} elements tends to enhance the geometry of the problem and, thus, it can reduce the positioning error. Also, the \ac{GDOP} depends strongly on the azimuth angle for small $\Nl$ and, consequently, on the \ac{MS} orientation. The same behavior can be seen in \cref{Fig:GDOPOEB} for the \ac{GDOP} related to the orientation, i.e., $\mathsf{GDOP}_{\amo}$. The main difference is that it is more harder to estimate the orientation angle with a small number of \ac{RIS} elements compared to the position estimation. In fact, the \ac{GDOP} can be interpreted as a mapping from the standard deviation of the thermal noise to the estimation error in terms of the \ac{PEB} and \ac{OEB}. For example, with $\Nl=4$, we have $\PEB \approx \,1.9\, \dbmo\, \sigma=5.7\, \sigma$, while $\OEB \approx 190\, \sigma$, from \eqref{eq:GDOPP} and  \eqref{eq:GDOPO}.
The reason is that the orientation estimation relies on the curvature of the wavefront in the near-field. Hence, for a larger number of elements, the effective size of the \ac{RIS} increases along with the Fraunhofer distance \cite{Balanis:05}.

\section{Conclusions}
\label{sec:conclusions}
In this paper, we propose an architecture for joint communication and \ac{MS} localization and orientation {estimation} in a \ac{RIS}-assisted environment. We derive the ultimate performance in terms of \ac{PEB} and \ac{OEB}, accounting for both near- and far-field propagation conditions. {The \ac{RIS} phases are  designed to maximize the \ac{SNR} towards the desired \ac{MS} for both communication and localization enhancement. Indeed, we obtained} that the \ac{RIS} with the proposed phase design can significantly increase the localization performance by focusing the incident spherical wavefront from the \ac{MS} toward the \ac{BS}. The proposed scheme, when compared to a conventional system without \ac{RIS}, can achieve up to two orders and one order of magnitude reduction in \ac{PEB} and \ac{OEB}, respectively. Also, the localization accuracy strongly depends on the considered geometry and the orientation of the \ac{MS}. The achieved results open the door towards the adoption of \acp{RIS} as an effective meaning for supporting mobile wireless localization {and, thus, to boost the communication performance}. A step forward will be the analysis of the localization performance limits in presence of multiple \acp{RIS}.

\section*{Acknowledgements}
Authors wish to thank D. Dardari for motivating this work and A. Abdelhady for the useful discussion about the optimization problem.


%

\begin{figure*}
\begin{equation}  
\mathbf{J}_{\pmo} = \left[\begin{array}{cccc cccc}  \nabla_{\xms} \rhobm & \nabla_{\xms} \thetabm &  \nabla_{\xms} \phibm & \nabla_{\xms} \taubm& 
\nabla_{\xms} \rhoblm & \nabla_{\xms} \thetalm & \nabla_{\xms} \philm & \nabla_{\xms} \taulm   \\
\nabla_{\yms} \rhobm & \nabla_{\yms} \thetabm &  \nabla_{\yms} \phibm & \nabla_{\yms} \taubm& \nabla_{\yms} \rhoblm & \nabla_{\yms} \thetalm & \nabla_{\yms} \philm & \nabla_{\yms} \taulm   \\
\nabla_{\zms} \rhobm & \nabla_{\zms} \thetabm &  \nabla_{\zms} \phibm & \nabla_{\zms} \taubm& \nabla_{\zms} \rhoblm & \nabla_{\zms} \thetalm & \nabla_{\zms} \phibm & \nabla_{\zms} \taulm   \\
\end{array} \right],
\label{eq:Jacobianmatrix}
\end{equation}
\end{figure*}
\appendices

\section{The Jacobian Matrix}
\label{app:Jacobian}

In this appendix, we report the elements of the Jacobian matrix for the \ac{CRLB} derivation in \eqref{eq:crb_general}. The Jacobian matrix of the mobile location is given by \eqref{eq:Jacobianmatrix} in the top of next page

%
where  for each  $\am\in\left\{\xmo, \ymo , \zmo \right\}$ and  $\mathsf{S}\in \left\{\mathsf{B}, \mathsf{R}\right\}$, the following relationships holds
\begin{align}
&\nabla_{\am}\, \tau_{\mathsf{SM}} = \frac{\nabla_{\am} \, d_{\mathsf{SM}}}{c}=\frac{1}{c} \, \frac{\am- a_{\mathsf{S}}}{d_{\mathsf{SM}}}, \\
&\nabla_{\am}\, \phi_{\mathsf{SM}} = \frac{1}{1+{\left(\frac{\yms-y_{\mathsf{S}}}{\xms-x_{\mathsf{S}}}\right)}^{2}}\, \nabla_{\am}\left( \frac{\yms-y_{\mathsf{S}}}{\xms-x_{\mathsf{S}}} \right), \label{eq:nablaphi}\\
&\nabla_{\am}\,\theta_{\mathsf{SM}}=\frac{1}{\sqrt{1-\left(\frac{\zms-z_{\mathsf{S}}}{d_{\mathsf{SM}}}\right)^2}}\, \nabla_{\am}\left( \frac{\zms-z_{\mathsf{S}}}{d_{\mathsf{SM}}}\right), \label{eq:nablatheta} \\
&\nabla_{\am}\, \rhobm =-\frac{\lambda}{4\,\pi} \, \frac{1}{\left(\dbmo\right)^2} \, \nabla_{\am} \left(\dbmo\right), \\
&\nabla_{\am}\, \rhoblm =- \frac{\lambda}{4\,\pi} \, \frac{1}{(\dlmo+\dblo)^2}  \frac{\am- a_{\mathsf{R}}}{\dlmo}.
\end{align}
After some manipulation, \eqref{eq:nablaphi}-\eqref{eq:nablatheta} can be simplified as
\begin{align}
&\begin{cases}
\nabla_{\xms}\, \phidm = 
- \frac{1}{\ddmo} \frac{\sin\phidm}{\cos\thetadm} \\
\nabla_{\yms}\,\phidm= 
\frac{1}{\ddmo} \frac{\cos\phidm}{\cos\thetadm} \\
\nabla_{\zms}\,\phidm   =0 
\end{cases}
&& \!\!\!
\begin{cases}
\nabla_{\xms}\,\thetadm =
- \frac{\sin\left( \thetadm \right) \cos\left(\phidm \right)}{\ddmo} \\
\nabla_{\yms}\, \thetadm =
- \frac{\sin\left( \thetadm \right) \sin\left(\phidm \right)}{\ddmo}\\
\nabla_{\zms}\, \thetadm=
\frac{\cos(\thetadm)}{ \ddmo}.
\end{cases}
\end{align}

\section{FIM Elements}
\label{app:FIMelements}
%

In order to derive the elements of the \ac{FIM} in \eqref{eq:FIMelements}, the derivatives of the mean received signal with respect to the parameters, i.e., $ {\partial \mu_{b}[n]}/{\partial \Gamma_{j}}$, should be derived for each $\Gamma_{j} \in \boldsymbol{\Gamma}$. Let us first rewrite \eqref{eq:mub} as
\begin{align}\label{eq:recsigsum3}
\mu_{b}[n]=\muBM + \muBLM,
\end{align}
with
\begin{align}
&\muBM  \triangleq \sqrt{P} \rhobm\, \sum_{m=1}^{\Nm} \mubm,  \\
& \muBLM  \triangleq \sqrt{P}\rhoblm \,\sum_{m=1}^{\Nm}\, \sum_{r=1}^{\Nl} \mublm.
\end{align}
The signal inside the summation is
\begin{align}
&\mubm  \triangleq 
\xm[n]    \operatorname{exp}\left(-j 2\pi \fn\, \left(\tilde{\tau}_{bm}+\tilde{\xi}_{\mathsf{BM}} +\etam \right)\right), 
\\
&\mublm \triangleq \xm[n] \, \omegal \nonumber \\
& \times \operatorname{exp}\left(-j2\,\pi \fn \left(\tilde{\tau}_{br}+\tilde{\tau}_{rm} + \etal +\tilde{\xi}_{\mathsf{BRM}} + \etam \right)\right), 
\end{align}
where we have the following definitions for the synchronous signalling: $\tilde{\xi}_{\mathsf{BM}}\triangleq \xibm $, $\tilde{\xi}_{\mathsf{BRM}} \triangleq \xibm $, $\tilde{\tau}_{bm}\triangleq\tau_{bm}$, $\tilde{\tau}_{rm}\triangleq\tau_{rm}$, and $\tilde{\tau}_{br}\triangleq\tau_{br}$; whereas for the asynchronous signalling it is: $\tilde{\xi}_{\mathsf{BM}} \triangleq \chibm/2\pi\fn$, $\tilde{\xi}_{\mathsf{BRM}} \triangleq \chiblm/2\pi\fn$, $\tilde{\tau}_{bm}\triangleq\Delta \tau_{bm}$, $\tilde{\tau}_{rm}=\Delta \tau_{rm}$, and $\tilde{\tau}_{br} \triangleq\Delta \tau_{br}$.

\paragraph*{Two-stage localization}
The derivatives for the indirect approach can be found now as 
\begin{align*}
& 
\nabla_{\rhobm}\, {\mu}_{b}[n]  = {\muBM}/{\rhobm},\,   
\nabla_{\rhoblm}\,  {\mu}_{b}[n] 
={\muBLM}/{\rhoblm}, \\
&\nabla_{\thetabm}\, {\mu}_{b}[n]=- j 2 \pi \fn\,\sqrt{P}\, \rhobm\, \sum_{m=1}^{\Nm}  \mubm\, \nabla_{\thetabm}\, \tbm, \\
%
& \nabla_{\thetalm} {\mu}_{b}[n]=
- j 2 \pi \fn\,\sqrt{P}\, \rhoblm\, \sum_{m=1}^{\Nm} \, \sum_{r=1}^{\Nl}  \mublm\,  \nabla_{\thetalm} \, \tlm,    \\
&\nabla_{\phibm} {\mu}_{b}[n]=
- j 2 \pi \fn\,\sqrt{P}\, \rhobm\, \sum_{m=1}^{\Nm}  \mubm\,  \nabla_{\phibm} \tbm,  \\
&  \nabla_{\philm} {\mu}_{b}[n]=
- j 2 \pi \fn\,\sqrt{P}\, \rhoblm\, \sum_{m=1}^{\Nm}  \, \sum_{r=1}^{\Nl}  \mublm\,  \nabla_{\philm}\, \tlm,   \\
&\nabla_{\taubm} {\mu}_{b}[n]=
- j 2 \pi \fn\,\sqrt{P}\, \rhobm\, \sum_{m=1}^{\Nm}  \mubm\,  \nabla_{\taubm} \tbm,
\\
&  \nabla_{\taulm}{\mu}_{b}[n]= 
-j 2 \pi \fn\, \sqrt{P}\, \rhoblm\, \sum_{m=1}^{\Nm} \, \sum_{r=1}^{\Nl}  \mublm\,  \nabla_{\taulm}\, \tlm,  
\end{align*}
As regards the derivatives with respect to the rotational angles of the mobile, we have $\forall\,  \amo \in \{\alphamo,\betamo,\gammamo\}$
\begin{align}
\nabla_{\amo} {\mu}_{b}[n]&= - j 2 \pi \fn\,\sqrt{P}\, \left( \rhobm\, \sum_{m=1}^{\Nm}  \mubm\,  \nabla_{\amo} \tilde{\tau}_{bm}
 \right. \nonumber \\
 &\left.\phantom{=}+ \rhoblm\, \sum_{m=1}^{\Nm}  \, \sum_{r=1}^{\Nl}  \mublm\,  \nabla_{\amo} \tilde{\tau}_{rm}\right). 
\end{align}

The derivatives of the \acp{TOA} with respect to the parameters can be written for each   $\mathsf{S}\in \left\{\mathsf{B}, \mathsf{R}\right\}$ and the corresponding antenna index $s\in\left\{b,r\right\}$ as 
\begin{align}
&\nabla_{\thetadm} \tdm
=\frac{\ddmo}{c\, \ddm}\left[ \left(\xm-\xd \right)\sin\thetadm \cos\phidm + \right. \nonumber\\
&\phantom{\nabla_{\thetadm}}\left. + \left(\ym-\yd \right)\sin\thetadm \sin\phidm -\left(\zm-\zd \right)\cos\thetadm \right], \\
&\nabla_{\phidm} \tdm
=\frac{\ddmo}{c\,\ddm}\left[\left(\xm-\xd \right)\cos\thetadm \sin\phidm + \right. \nonumber\\
&\phantom{\nabla_{\phidm}}\left. - \left(\ym-\yd \right)\cos\thetadm \cos\phidm \right], \\
&\nabla_{\amo} \tdm=\frac{1}{c\, \ddm}\left[\xm \,\nabla_{\amo} \xm+ \ym\, \nabla_{\amo} \ym+\zm\, \nabla_{\amo} \zm  \right. \nonumber \\
&\phantom{\nabla_{\amo}}- \left. \left(\xd \,\nabla_{\amo} \xm+ \yd\, \nabla_{\amo} \ym+\zd\, \nabla_{\amo} \zm +
\right. \right. \nonumber\\
&\phantom{\nabla_{\amo}}\left. \left. +\ddmo \left( \nabla_{\amo} \xm \cos\thetabm \cos\phibm  + \right. \right. \right. \nonumber\\
&\phantom{\nabla_{\amo}}\left. \left. \left.
+ \nabla_{\amo} \ym \cos\thetabm \sin\phibm +\nabla_{\amo} \zm \sin\thetabm \right) \right) \right],
\end{align}
\begin{align}
&\nabla_{\taudm} \tdm = \frac{1}{\ddm} \left[\ddmo- G_{dm}^{(2)} \right], \\
&\nabla_{\amo} \Delta\tdm =\nabla_{\amo} \tdm, \\
&\nabla_{\thetadm} \Delta\tdm= \nabla_{\thetadm} \ddm,\nonumber\\
&\nabla_{\phidm} \Delta\tdm =\nabla_{\phidm} \ddm, \\
&\nabla_{\taudm} \Delta \tdm= \nabla_{\taudm}  \tdm - 1,
\end{align}
where
\begin{align*}
\nabla_{\alphamo} \xm &=-\sinam\cosbm \xmi + \left[-\sinam\sinbm\singm-\cosam\cosgm\right]\, \ymi \nonumber \\
&\phantom{=} +\left[\cosam \singm - \sinam \sinbm \cosgm   \right] \zmi,  \\
\nabla_{\betamo} \xm &=-\cosam \sinbm \xmi + \cosam \cosbm \singm\, \ymi +\cosam\cosbm\cosgm  \zmi,  \\
\nabla_{\gammamo} \xm &= \left[\cosam\sinbm\cosgm+\sinam\singm \right]\, \ymi+ \nonumber \\
&\phantom{=}+\left[\sinam\cosgm - \cosam\sinbm\singm   \right] \zmi,
\end{align*}
\begin{align*}
\nabla_{\alphamo} \ym &=\cosam\cosbm \xmi + \left[-\cosgm\sinam+ \cosam\sinbm\singm \right]\, \ymi + \nonumber \\
&\phantom{=}+\left[\cosgm\cosam\sinbm +\sinam\singm \right] \zmi, \\
\nabla_{\betamo} \ym &=-\sinam\sinbm \xmi +  \sinam\cosbm\singm \, \ymi  +\cosgm\sinam\cosbm \zmi, \\
\nabla_{\gammamo} \ym &= \left[-\singm\cosam+ \sinam\sinbm\cosgm \right]\, \ymi + \nonumber \\
&\phantom{=}-\left[\singm\sinam\sinbm +\cosam\cosgm \right] \zmi, \\
\nabla_{\alphamo} \zm &=0,  \\
\nabla_{\betamo} \zm &=-\cosbm \xmi - \sinbm\singm\, \ymi - \sinbm\cosgm \zmi, \\
\nabla_{\gammamo} \zm &= \cosbm\cosgm\, \ymi - \cosbm\singm \zmi,
\end{align*}
where $c_{x_{\mathsf{M}}}\triangleq \cos\left(x_{\mathsf{M}}\right)$ and $s_{x_{\mathsf{M}}}\triangleq \sin\left(x_{\mathsf{M}}\right)$.

\paragraph*{Direct localization}
When a direct localization approach is used, the signal can be rewritten as
\begin{align}
 \!\!\mu_b &=  \sum_{m=1}^{\Nm} \left(f_{bm}\left(\rhobm, d_{bm}\right)+  \sum_{r=1}^{\Nl} g_{brm} \left(\rhoblm, \dlm \right) \right),  \label{eq:recsigsum2}
\end{align}
where  $f_{bm}$ and $g_{brm}$ are two non-linear functions depending on the parameters to be estimated,\footnote{Generally, the optimal design of \ac{RIS} phases can depend on the \ac{MS} location and orientation. For convenience, such a dependence is neglected in \eqref{eq:recsigsum2}.} defined as\footnote{Here we have dropped the synchronization mismatches and array errors.}
\begin{align}
&f_{bm}\left(\rhobm, \dbm \right)\!\triangleq\! \sqrt{P} \! \,\xm[n]\rhobm(\pmo) \operatorname{exp}\!\left(-j \,2\pi  \frac{\fn}{c}\,  \tilde{d}_{bm}  \right)\!, 
\end{align}
where $\tilde{d}_{bm} = \tilde{d}_{bm} \left( \pmo,\amo\right) \triangleq  \dbm$ and $\tilde{d}_{bm} \triangleq \Delta \dbm$ in synchronous and asynchronous cases, respectively, and
\begin{align}
g_{brm} \left(\rhoblm, \dlm  \right) &\triangleq \sqrt{P} \,\xm[n] \rhoblm(\pmo)\,\, \omegal \cdot \nonumber \\
&\times \operatorname{exp}\left(-j\,2\,\pi \frac{\fn}{c} \left(\tilde{d}_{br}+\tilde{d}_{rm} \right) \right), 
\end{align}
where $\tilde{d}_{rm} = \tilde{d}_{rm} \left( \pmo,\amo\right) \triangleq  \dlm$  and $\tilde{d}_{rm} \triangleq \Delta \dlm$ in synchronous scheme; whereas $\tilde{d}_{br} = \dbl$ and $\tilde{d}_{br} = \Delta \dbl$ in asynchronous scheme.

The gradient vector with respect to the position, $\pmo$, and orientation, $\amo$,  can be written from  \eqref{eq:recsigsum2}  as 
\begin{align}\label{eq:gradientvecgb}
\nabla_{\pmo}\left( \mu_b \right) &= \left[ \nabla_{\xms}\, \mu_{b},\, \nabla_{\yms}\, \mu_{b},\, \nabla_{\zms} \, \mu_{b} \right] = \nonumber \\
&=\sum_{m=1}^{\Nm}\,  \nabla_{\pmo} f_{bm} +  \sum_{r=1}^{\Nl} \nabla_{\pmo} g_{brm}, \\
\nabla_{\amo}\left( \mu_b \right) &= \left[ \nabla_{\alphamo}\, \mu_{b},\, \nabla_{\betamo}\, \mu_{b},\, \nabla_{\gammamo}\, \mu_{b} \right] = \nonumber \\
&= \sum_{m=1}^{\Nm}\,  \nabla_{\amo} f_{bm} +  \sum_{r=1}^{\Nl} \nabla_{\amo} g_{brm},
\end{align}
where for the direct path we have
\begin{align}
\nabla_{\pmo} f_{bm} &= 
\sqrt{P} \, x_m \, \left[\nabla_{\dbmo} \rhobm \nabla_{\pmo} \dbmo   + \right. \nonumber \\
&\left.\phantom{=} - j\, 2\, \pi \, \frac{\fn}{c} \rhobm \nabla_{\pmo} \tilde{d}_{bm} \right]\, e^{-j\, 2\, \pi \, \frac{\fn}{c} \, \tilde{d}_{bm}},
\nonumber \\
\nabla_{\amo} f_{bm} &= 
-j\, 2\, \pi \, \frac{\fn}{c} \sqrt{P}   \, \rhobm \, x_m  \, e^{-j\, 2\, \pi \, \frac{\fn}{c} \,   \nabla_{\amo} \tilde{d}_{bm}},  
\end{align}
while for the \ac{RIS}-relayed path it is
\begin{align}
\nabla_{\pmo}\, g_{brm} &= 
\sqrt{P} \, x_m \, \omegal \left[\nabla_{\dlm} \rhoblm \nabla_{\pmo} \dlmo + \right. \nonumber \\
&\left. \phantom{=}-j\, 2\, \pi \, \frac{\fn}{c}  \rhoblm \nabla_{\pmo} \tilde{d}_{rm} \right] e^{-j\, 2\, \pi \, \frac{\fn}{c} \, \left(\tilde{d}_{rm} +\tilde{d}_{br}\right)}, 
\nonumber\\ 
\nabla_{\amo}\, g_{brm} &= 
-j\, 2\, \pi \, \frac{\fn}{c}  \sqrt{P} \, \rhoblm\,  x_m \, \omegal \, e^{-j\, 2\, \pi \, \frac{\fn}{c} \, \left(\tilde{d}_{rm} +\tilde{d}_{br}\right)}\nonumber \\
&\phantom{=}\times \nabla_{\amo} \tilde{d}_{rm}, 
\end{align}

The derivatives of the path-loss amplitudes with respect to the distances between array centers are
\begin{align}
&\nabla_{\dbmo}\,\, \rhobm = -\frac{\lambda}{4 \pi \dbmo^2}, && \nabla_{\dlmo}\,\, \rhoblm = -\frac{\lambda}{4 \pi (\dblo+ \dlmo)^2}.
\end{align}
By denoting with $\am\in \left\{ \xmo, \ymo, \zmo \right\}$, $\mathsf{S}\in \left\{\mathsf{B}, \mathsf{R}\right\}$ and $s\in \left\{b, r\right\}$, we can obtain
\begin{align} \label{eq:difdbm}
&\nabla_{\am}\, \Delta \ddm = \nabla_{\am}\,  \ddm - \nabla_{\am}\, \ddmo, \\
&\nabla_{\am}\, \ddm =  \nonumber \\
&\phantom{\nabla} =\frac{1}{2 \, \ddm}   \nabla_{\am}\left(\dm^2 + d_s^2 + \ddmo^2 -2\left(\mathsf{G}_{sm}^{(1) }+ \ddmo \,\mathsf{G}_{sm}^{(2)}\right)\right),   \nonumber \\
&\phantom{\nabla}= \frac{1}{\ddm} \left( \ddmo  \nabla_{\am}\ddmo - \nabla_{\am}\ddmo \,\gtdm  -  \ddmo\, \nabla_{\am}\gtdm \right) ,
\end{align}
where
$\nabla_{\am}\, \ddmo(\am)= \frac{\am-a_{\mathsf{S}}}{\ddmo}$ and where 
\begin{align}\label{eq:Gtdm2}
\nabla_{\am}\,\gtdm &=  -\left(\xm-\xd \right)\, \sin(\thetadm) \, \cos(\phidm) \, \nabla_{\am}\, \thetadm \nonumber \\
&\phantom{=}-  \left(\xm-\xd \right) \, \cos(\thetadm) \, \sin(\phidm) \, \nabla_{\am}\, \phidm   \nonumber \\
&\phantom{=}- \left(\ym-\yd \right) \, \sin(\thetadm) \, \sin(\phidm) \, \nabla_{\am}\, \thetadm \nonumber \\
&\phantom{=}+  \left(\ym-\yd \right) \, \cos(\thetadm) \, \cos(\phidm) \, \nabla_{\am}\, \phidm \nonumber \\
&\phantom{=}+\left(\zm-\zd \right) \, \cos(\thetadm) \,  \nabla_{\am} \thetadm,
\end{align}
with $ \nabla_{\am}\, \phidm$ and $\nabla_{\am}\, \thetadm$ as in Appendix~\ref{app:Jacobian}.
Thus, the derivatives in \eqref{eq:Gtdm2} becomes
\begin{align}
&\nabla_{\xms}\, \gtdm=  
\frac{\xm-\xd}{\ddmo} \, \left[\sin^2\thetadm \,  \cos^2\phidm+ \sin^2\phidm \right]  \nonumber \\
&\phantom{\nabla_{\xms}}+ \frac{\ym \!-\!\yd}{\ddmo} \left[\sin^2\thetadm \, \cos\phidm \sin\phidm  -  \sin\phidm \cos\phidm \right]  \nonumber \\
&\phantom{\nabla_{\xms}}-\frac{\zm-\zd}{\ddmo}\sin\thetadm \, \cos\phidm \,\cos\thetadm,\\
&\nabla_{\mathsf{\yms}}\, \gtdm =  \frac{\xm-\xd}{\ddmo} \left[\sin^{2}\thetadm\, \sin\phidm \, \cos\phidm  \right. \nonumber \\
&\phantom{\nabla_{\mathsf{\yms}}} \left. - \sin\phidm \, \cos\phidm \right] +\frac{\ym-\yd}{\ddmo} \left[\sin^{2}\thetadm \, \sin^{2}\phidm \, \right.  \nonumber\\
&\phantom{\nabla_{\mathsf{\yms}}} \left. +\cos^{2}\phidm\right]-\frac{\zm-\zd}{\ddmo} \left[ \sin\thetadm \, \cos\thetadm \, \sin\phidm \right], \\
&\nabla_{\mathsf{\zms}}\, \gtdm = - \frac{\xm-\xd}{\dbmo} \sin\thetadm \, \cos\thetadm\, \cos\phidm   \nonumber\\
&\phantom{\nabla_{\mathsf{\zms}}}+\frac{\ym-\yd}{\ddmo} \left[\sin\thetadm \, \cos\thetadm \,  \sin\phidm\right] \nonumber \nonumber \\
&\phantom{\nabla_{\mathsf{\zms}}}+\frac{\zm-\zd}{\ddmo} \cos^{2}\thetadm.
\end{align}

\bibliographystyle{IEEEtran}
\bibliography{IEEEabrv,LISBIB}

\end{document}